\documentclass[AMA,STIX1COL]{WileyNJD-v2}
\usepackage{moreverb}
\usepackage{lscape}
\usepackage{multirow}
\usepackage[dvips,colorlinks,bookmarksopen,bookmarksnumbered,citecolor=red,urlcolor=red]{}

\def\bea{\begin{eqnarray*}}
\def\eea{\end{eqnarray*}}
\def\bean{\begin{eqnarray}}
\def\eean{\end{eqnarray}}
\def\be{\begin{equation}}
\def\ee{\end{equation}}
\def\Cov{\mathop{\rm Cov}\nolimits}
\def\Corr{\mathop{\rm Corr}\nolimits}
\def\Var{\mathop{\rm Var}\nolimits}
\def\Ex{\mathop{\rm E}\nolimits}
\def\Bl{\Bigl}
\def\Br{\Bigr}
\def\bi{\begin{itemize}}
\def\ei{\end{itemize}}
\newcommand\BibTeX{{\rmfamily B\kern-.05em \textsc{i\kern-.025em b}\kern-.08em
T\kern-.1667em\lower.7ex\hbox{E}\kern-.125emX}}

\newcommand\pd{{t_1}}
\newcommand\pfs{{\rm PFS}}
\newcommand\os{{\rm OS}}
\newcommand{\dif}{{\rm \;d}}
\newcommand{\sos}{S_{OS}}
\newcommand{\spfs}{S_{PFS}}

\articletype{Article Type}%

\received{24 October 2018}

\begin{document}

\title{Joint modelling of progression-free and overall survival and
  computation of correlation measures}

\author[1]{Matthias Meller*}

\author[2]{Jan Beyersmann}

\author[3]{Kaspar Rufibach}

\authormark{Meller \textsc{et al}}

\address[1, 3]{\orgdiv{Department of Biostatistics}, \orgname{F. Hoffmann-La Roche Ltd},
  \orgaddress{\state{Basel}, \country{Switzerland}}}

\address[2]{\orgdiv{Institute of Statistics}, \orgname{Ulm University}, \orgaddress{\state{Helmholtzstrasse 20, 89081 Ulm}}, \country{Germany}}


\corres{*Matthias Meller. \email{meller.matthias@gmail.com}}


\abstract[Summary]{In this paper we derive the joint distribution of
  progression-free and overall survival as a function of transition
  probabilities in a multistate model. No assumptions on copulae
    or latent event times are needed and the model is allowed to be
    non-Markov. From the joint distribution, statistics of interest can then
  be computed. As an example, we provide closed formulas and statistical
  inference for Pearson's correlation coefficient between progression-free and
  overall survival in a parametric framework. The example is
    inspired by recent approaches to quantify the dependence between
    progression-free survival, a common primary outcome in phase III trials in
    oncology, and overall survival. We complement these approaches by
    providing methods of statistical inference while at the same time working
    within a much more parsimonious modelling framework. Our approach is
    completely general and can be applied to other measures of dependence.  We
    also discuss extensions to nonparametric inference.
  Our analytical results are illustrated using a large randomized clinical
  trial in breast cancer.}

\keywords{event history analysis, randomized clinical trials, multistate model, illness-death model}

\jnlcitation{\cname{%
\author{Meller M},
\author{Beyersmann J}, and
\author{Rufibach K}} (\cyear{xxxx}),
\ctitle{Joint modelling of progression-free and overall survival and computation of correlation measures}, \cjournal{Stat Med}, \cvol{xxxx;xx:x--x}.}

\maketitle


\section{Introduction}
\label{intro}
An established way to evaluate performance of therapies in clinical trials are
time-to-event endpoints. While overall survival (OS) remains the gold standard
for demonstrating clinical benefit, especially in oncology, alternative
endpoints such as progression-free survival (PFS), are also accepted by Health
Authorities\cite{fda_endpoints, pazdur_08}. PFS is not
only considered a surrogate for OS, but may provide clinical benefit by
itself, e.g. by delaying symptoms or subsequent therapies
\cite{george_16}. Further advantages of the use of PFS are shorter trial
durations and the fact that, as opposed to OS, PFS is not confounded by the
use of later line, i.e. post-progression, therapies, and PFS is less
vulnerable to competing causes of death than OS\cite{saad_16}.

Methodology has been developed to assess surrogacy of one endpoint for
another\cite{burzykowski_01} and these methodologies have been
applied to a wide range of indications\cite{buyse_16}. One important aspect
in the assessment of surrogacy is to quantify the correlation between the
surrogate and the real endpoint\cite{prasad2015}. 
This aspect has received quite some attention in the literature lately\cite{fleischer2009statistical, li2015weibull}.
These two papers consider Pearson's correlation coefficient and
rely on an illness-death model without recovery (IDM) to model the association
between PFS and OS. To specify the underlying statistical model for the
likelihood function for parametric estimation, they use a latent failure time
approach. One characteristic of the latent failure time approach
  is that it allows for PFS events after OS, see Section~\ref{subsec:latent}.
Within this model formulation, Fleischer et al.\cite{fleischer2009statistical} then derive
closed formulas for the survival functions $\spfs$ and $\sos$ for PFS and OS
as well as the correlation coefficient $\Corr(\pfs, \os)$ by assuming an
exponential distribution for all transition intensities in the underlying
multistate model. Li and Zhang\cite{li2015weibull} generalized these results to
Weibull transition intensities, where closed formulas are provided imposing
the assumption of equal Weibull shape parameter for all three transition
intensities.

Notably, while {\color{blue}in} Fleischer et
al.\cite{fleischer2009statistical} and Li and Zhang\cite{li2015weibull} point
estimates are provided for the quantities of interest based on the made
parametric assumptions, they do not provide a discussion of statistical
inference, e.g. how to derive pointwise confidence intervals for the
distribution functions at a milestone time or for the correlation coefficient
directly. We further discuss the contributions of these papers in detail in
Section~\ref{sec:inf}. One aim of this paper is to provide for such methods of
statistical inference while ensuring that PFS does not exceed OS.


Finally, in a Bayesian framework the joint distribution of PFS and OS can be modelled based on a Gaussian copula\cite{fu_13}. Inference is based on a likelihood function that relies on the latent failure time approach, i.e. derives each patient's contribution to the likelihood based on his or her censoring pattern.

We extend the existing body of work in multiple directions. To
  begin, we model PFS and OS as arising from an IDM without recovery. The only
  assumption will be that the model is smooth in the sense that transition
  intensities between all states of the model exist. No assumption about
  latent times or copulas will be made and the model is allowed to be
  non-Markov. Because the transition probabilities provide a full description
  of the IDM, we first recollect closed formulas for the transition
probabilities in an IDM process~$X$,
  following Hougaard~\cite{hougaard_00}. We derive these probabilities for $X$ a
general, not necessarily Markov, process. The implication of $X$ being
non-Markov for the considered IDM is that the hazard for the transition from
the diseased into the death state, $\alpha_{12}(t; \pd)$, will depend on the
time $t$ since time origin {\it and} on the time $\pd$ of entry into the
diseased state. While conceptually offering more modelling flexibility, this
generalization complicates derivation of closed formulas for the quantities of
interest, since one additionally needs to integrate out over the distribution
of $\pd$. There are multiple ways how $\alpha_{12}(t; \pd)$ can depend on
$\pd$, e.g. as function of $t$ and $\pd$ separately or via the duration $t -
\pd$ only. Naming of these specific assumptions has not been unambiguously in
the literature and we provide a discussion of the terminology. 

Based on the transition probabilities, we derive closed formulas not only for
$\spfs$, $\sos$, and $\Corr(\pfs, \os)$, but also for the {\it joint}
distribution of $\pfs$ and $\os$. It is worth noting that the latter entirely
specifies the association between PFS and OS based on the assumed IDM model
alone, i.e. without having to make additional assumptions, e.g. about a copula
function as in Burzykowski et al.\cite{burzykowski_01}. 
For all these quantities, and under parametric assumptions on the
  transition hazards, statistical inference can be performed either via
standard likelihood theory together with the $\delta$-rule or bootstrap for
parametric models. Note that, unlike in Li and Zhang\cite{li2015weibull}, with the IDM
formulation there is no need to restrict the Weibull shape parameter to
coincide between all three transition intensities.

There is a rich theory allowing for nonparametric estimation of transition
hazards and subsequently probabilities, see e.g. Hougaard \cite{hougaard_00} or
Aalen et al.\cite{aalen_08}. Through simple plug-in of these estimates we also get
nonparametric estimates for the above quantities, thereby extending the existing purely
parametric approaches\cite{fleischer2009statistical, li2015weibull}. Inference for these models is possible via
bootstrap. Estimating transition probabilities nonparametrically based on
possibly right-censored data adds one complication though: to estimate a
correlation coefficient we rely on expectations and variances with respect to
$\spfs$ and $\sos$, i.e. we either need estimates of these survival
functions down to 0 or we restrict estimation of expectations to some maximum
time.

It is well-known that the latent failure formulation of competing risk or
multistate models in general suffers from identifiability and interpretability
issues. We discuss these in more detail and shed some light on how the latent
times, if they are introduced, relate to the transition hazards from an
IDM. In contrast to using latent times, copulas may be applied to
  modelling the dependence of PFS and OS without assuming such a latent
  structure. However, typical copula models consider the dependence of general
  multivariate survival times, such as the lifetimes of twins, and neither
  allow for PFS $=$ OS with positive probability nor ensure the natural
  PFS~$\le$~OS. These two aspects are accounted for in a multistate model,
  which makes the latter framework a natural choice for joint modelling of PFS
  and
  OS.

  Lately, Weber and Titman\cite{weber_18} also pointed out the mentioned
  deficiencies of copulas and the latent-failure time model used in
  \cite{fleischer2009statistical, li2015weibull} to model the association
  between PFS and OS. They also advocate an IDM, but, in contrast
    to our paper, focus on demonstrating the shortcomings of copula modelling
    for PFS and OS using Kendall's $\tau$, also using the likelihood approach
    of Li and Zhang\cite{li2015weibull}.

We would like to emphasize that our results are neither specific
to the modelling of PFS and OS in oncology nor to investigating their correlation as one choice of measuring dependence, but are
applicable to any IDM.

In Section~\ref{sec:msm} we introduce a multistate model that allows to
explicitly derive the joint distribution of PFS and OS under virtually no
assumptions and we show how this can be used to derive a formula for Pearson's
correlation coefficient. Furthermore, we outline that patient trajectories for
PFS and OS from our model can easily be simulated. Section~\ref{sec:fleischli}
revisits existing approaches using the latent failure time approach and
discusses how our proposal allows to extend these methods using more
parsimonious assumptions. How to make inference in our proposed setup is
described in Section~\ref{sec:inf} while Section~\ref{sec:estcorr} reports on
a small simulation study for the correlation coefficient; additional simulation results are provided in the online supplement. In
Section~\ref{sec:example} we then illustrate the developed methodology using a
large Phase 3 randomized oncology clinical trial and compare it to previously
proposed approaches. We conclude with a discussion in
Section~\ref{sec:discussion}.

\section{A multistate model for PFS and OS}
\label{sec:msm} We introduce the general model in Section~\ref{sub:msm:1}
along with an algorithm to generate data from the model, which can, e.g., be
used for numerical approximations. The model may or may not be
 time-inhomogeneous Markov, and violations of the Markov assumption are summarized
  in Section~\ref{sec:assNM}. Section~\ref{subsec:latent}
discusses advantages of our approach over latent failure time and
  over copula modelling. We also discuss how to derive the model specification
  within a latent times approach as in Fleischer et al.\cite{fleischer2009statistical} and Li and Zhang\cite{li2015weibull}
  based on our more parsimonious model. Both transition
probabilities of the multistate model and the joint distribution of PFS and OS
are in Section~\ref{sec:jointdist}. Based on the joint distribution,
dependency measures of PFS and OS can be expressed, and we exemplarily
consider the correlation in Section~\ref{sub:msm:last}. We
  reiterate that our derivations hold for a general IDM without recovery. The
  model is not required to be Markov, the transition intensities may be
  time-dependent and their parametric specification may depart from those in Fleischer et al.\cite{fleischer2009statistical} and Li and Zhang\cite{li2015weibull}. Dependence measures other
  than Pearson's correlation coefficient can be expressed, e.g. transition intensities can be plugged into (10) in Weber and Titman\cite{weber_18} to receive Kendall's $\tau$. 

\subsection{The general model and generation of PFS-OS trajectories}\label{sub:msm:1}
\begin{figure}[h!]
\begin{center}
\includegraphics[trim=0cm 21cm 0cm 4.5cm,clip]{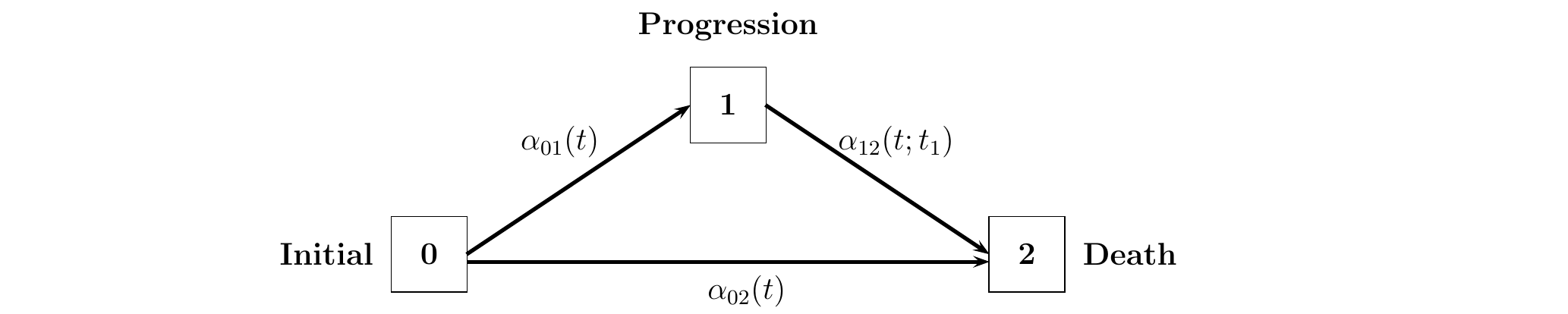}
\end{center}
\caption{Multistate model for PFS and OS: PFS is the waiting time in the initial state~$0$, OS is the waiting time until absorption in state~$2$.}
\label{fig:msm}
\end{figure}

We jointly model progression and death in an `illness-death' multistate model
\cite{Ande:Keid:mult:2002,bam}. The model is illustrated in
Figure~\ref{fig:msm}.

%
%
%
%
%
%
%
%
%
%

Let $X(t)$ denote the state occupied at time~$t$, $t\ge 0$, $X(t)\in
{\cal{S}} : = \{0,1,2\}$. We assume that all individuals are in the initial state~$0$ at
time~$0$, $P(X(0)=0)=1$, typically upon randomization to treatment. At the
time of diagnosis of progression, patients make a~$0\to 1$ transition, and
subsequent death is modelled as a~$1\to 2$ transition. Patients who die
without prior progression diagnosis make a direct $0\to 2$ transition into
the absorbing state. Progression-free survival is the waiting time in the
initial state~$0$,
\bea
  \pfs &=& \inf\{t\,:\, X(t) \neq 0\},   \label{eq:pfs}
\eea
and overall survival is the time until reaching the death state~$2$,
\bea
  \label{eq:os}
  \os &=& \inf\{t\,:\, X(t) = 2\}.
\eea
Patients with a diagnosed progression have~$\pfs < \os$, while patients who
make a direct $0\to 2$ transition have~$\pfs = \os$. At this stage, it is
important to note that no assumption about any dependence or independence of
PFS and OS has entered so far, other than the natural~$\pfs \le \os$. That
is, the only assumption we have made so far is that there is no progression
\emph{after} death. We will assume that the multistate model is sufficiently
smooth in that the following intensities exist,
\be
  \label{eq:csh}
  \alpha_{0j}(t) = \lim_{\Delta t \searrow 0} \frac{P(\pfs\in [t, t+\Delta
    t), X(\pfs)=j\,|\, \pfs \ge t)}{\Delta t},\, j=1,2,
\ee
and, for $\pd < t$,
\bean
  \alpha_{12}(t;\pd) & = & \lim_{\Delta t \searrow 0} \frac{P(X(t +\Delta t)=2\,|\, X(t-)=1, \pfs = \pd)}{\Delta t} \label{eq:csh12}\\
  {} & = & \lim_{\Delta t \searrow 0} \frac{P(\os - \pfs \in [t-\pd, t-\pd + \Delta t)\,|\, \os \ge t, \pfs = \pd)}{\Delta t}.\nonumber
\eean

These definitions are generally valid, also for $X(t)$ a non-Markov
process. In Section~\ref{sec:assNM} we discuss the classification of assumptions on $X$ and the implications for the definition of $\alpha_{12}$ in more detail.

The waiting times in the initial state and in the intermediate
  state in the model of Figure~\ref{fig:msm} have survival distributions
\bean \spfs(t) \ = \ P(\pfs > t) \ = \exp\left(-\int_0^t \alpha_{01}(u) +
  \alpha_{02}(u) \dif u\right) \label{eq:Spfs} \eean and,
  conditional on progression at time~$t_1$,
\bean
    P(\os - \pfs > v\, |\, \pfs=\pd < \os) &=& \exp\left(-\int_\pd^{\pd + v} \alpha_{12}(u;\pd)\dif u\right). \label{eq:condS}
\eean

Noting that $\pfs < \os$ given $\pfs = \pd$ with probability
$\alpha_{01}(\pd)/(\alpha_{01}(\pd) + \alpha_{02}(\pd))$
\cite{beyersmann08:_simul}, these formulae can be used to simulate
PFS-OS-trajectories \cite{bam} without resorting to latent
  times. One first simulates a PFS time using Equation~\eqref{eq:Spfs}, say
  $\pfs=\pd$. In a second step, one decides on $\pfs < \os$, if a binomial
  experiment with `success' probability $\alpha_{01}(\pd)/(\alpha_{01}(\pd) +
  \alpha_{02}(\pd))$ yields `success'. In this case, the residual time until
  OS is simulated from Equation~\eqref{eq:condS}. Otherwise, $\pfs=\os$. This
simulation approach is advantageous for numerical
approximations. Below, we will find that some expressions may be rather
tedious in the general non-Markov case, but they may be simply approximated by
their empirical counterparts after having generated a large number of
trajectories.

\subsection{Assumptions for non-Markov models}
\label{sec:assNM}

In a non-Markov IDM, i.e., without
the possibility of $1\to 0$ transitions, and common initial state 0, the $1
\to 2$ hazard $\alpha_{12}(t;\pd)$ will depend on time $t$ since time origin
and on time $\pd$ of entry into state 1, see Chapter 5 in Hougaard\cite{hougaard_00}
or Chapter 12 in Beyersmann et al.\cite{bam}. Since all transitions other than $1 \to 2$,
i.e. $0 \to 1$ and $0 \to 2$, are rooted in 0, making the transition intensity
$\alpha_{12}(t;\pd)$ depend on $\pd$ summarizes the entire ``past'' prior to
the $1 \to 2$ transition.

Now, classification of non-Markov models does not seem to be entirely
consistent in the literature. A model for which
$\alpha_{12}(t;\pd)$ depends only on the time
since progression, $\alpha_{12}(t;\pd) = \alpha_{12}(t - \pd)$ for
all~$\pd<t$, goes by the name {\it semi-Markov} in Hougaard (Table
6.1)\cite{hougaard_00} and Beyersmann et al. (p. 229)\cite{bam}, or {\it clock reset} in Putter et al.\cite{putter_07}. The
latter refers to the fact that the clock is re-set for the death hazard for
patients that progressed. Andersen and Pohar Perme\cite{andersen_08} and Andersen\cite{andersen_17} refer to
this model as {\it homogeneous semi-Markov} and to a model that depends on
both $t$ and $t - t_1$, i. e. $\alpha_{12}(t;\pd) = \alpha_{12}(t; t - \pd)$,
as {\it semi-Markov}. Equivalently, the general model,
$\alpha_{12}(t;\pd) = \alpha_{12}(t;\pd)$, is denoted {\it general Markov
  extension model}\cite{hougaard_00} and simply {\it
  non-Markov model}\cite{bam}.

If the intensities are time-constant, $\alpha_{ij}(t) = \alpha_{ij}$, the model is (time-) homogeneous Markov. Note that $X(t)$ being homogeneous Markov always implies exponential transition intensities and vice-versa, see p. 465 in Aalen et al.\cite{aalen_08} or p. 31 in Beyersmann et al.\cite{bam}.




\subsection{Advantages of the multistate modelling approach}
\label{subsec:latent}

In the literature, it is not uncommon that PFS and OS are
  modelled as arising from latent failure times
  \cite{Buys:Pied:on:1996,goldman_08,fleischer2009statistical,heng_11,crowther2013simulating,li2015weibull,xia_16,nomura_17}. Another
  approach is to use copulas to capture the dependence between PFS and OS
  \cite{burzykowski_01,weir2006statistical,buyse2007progression,fu_13,takeshi2017}.
  Neither latent times nor copulas have been used in our general model of
  Section~\ref{sub:msm:1}, and this is a conceptual advantage of the
  multistate modelling framework. To begin, copulas are a convenient method to
  model general bivariate survival data, e.g., lifetimes of twins, but the
  structure at hand is more specific: Firstly, PFS is always less than or
  equal to OS, but copulas do not place such a restriction on the pair of
  event times. This is meaningful for general bivariate survival data, but not
  for the ordered pair $(\mbox{PFS}, \mbox{OS})$. Secondly, PFS~$=$~OS with
  positive probability. However, in a copula model with continuous `marginal'
  survival functions for PFS and OS, this will happen with probability zero
  only. Again, this is meaningful for general bivariate survival data, but not
  for $(\mbox{PFS}, \mbox{OS})$, where PFS~$=$~OS simply encodes that the
  waiting time in the intial state equals the waiting time until the terminal
  state.

  For discussion of the latent failure time model, assume that there are
latent, non-negative random variables~$L_1$ and~$L_2$ such that progression
occurs, if $L_1 < L_2$ (and then $\pfs = L_1$ and $\os=L_2$). If, however,
$L_2 \le L_1$, the patient dies without prior progression (and then $\os =
L_2$). There are two problems with this approach, the first of which may
appear to be the more serious one, because it seemingly complicates estimating
the correlation or the joint distribution of PFS and OS, but it really is the
second problem that stands out: Firstly, it is well known from the competing
risks literature\cite{Kalb:Pren:stat:2002} that one cannot tell from the
observable data whether or not $L_1$ and $L_2$ are independent. However, this
problem is insolubly tied to assuming the existence of latent times. The
problem simply does not exist within the multistate framework, which is why
Aalen called the non-identifiable dependence structure of~$L_1$ and~$L_2$ an
`artificial problem' \cite{Aale:dyna:1987}. More important is a second
problem, namely whether there is any medical insight to be gained by assuming
latent times \cite{Aale:dyna:1987}, and we will demonstrate below that one
easily expresses the joint distribution of PFS and OS in terms of the
multistate model without such a latent structure. Chiang
\cite{chiang1991competing} summarizes the problem by stating that the latent
failure time structure requires an impossible sampling space: $L_1 > L_2$
implies that there is progression \emph{after} death, which is an awkward
idea, to say the least.

\emph{If} one assumes the additional latent failure time structure, the
connection to the multistate model is as follows. Because of the
non-identifiable dependence structure of~$L_1$ and~$L_2$, these are typically
assumed to be independent with marginal survival functions $P(L_j>t) =
\exp(-\int_0^t\alpha_{L_j}(u)\dif u)$. It is then well known from the
competing risks literature \cite{Kalb:Pren:stat:2002} that $\alpha_{L_j}(t) =
\alpha_{0j}(t)$. We also have for $t_1 < t$ that
\begin{eqnarray*}
P(L_2 - L_1 \in [t-\pd, t-\pd + \Delta
    t)\,|\, L_1 = t_1, L_2 \ge t) & = & P(L_2 \in [t, t + \Delta
  t)\,|\, L_1 = t_1, L_2 \ge t)\\
  {} & = &  P(L_2 \in [t, t + \Delta
  t)\,|\, L_2 \ge t),
\end{eqnarray*}
such that $\alpha_{12}(t;\pd) = \alpha_{L_2}(t)$. Obviously, this
  is not a convincing property, namely that progression has no effect on the
  hazard of death, such that authors tend to introduce a third time,
  say~$L_3$, conditional on $L_1 < L_2$
  \citep{fleischer2009statistical,li2015weibull}, which models the residual
  time until death after progression. One downside of this approach is that only
  $L_1$ and $L_3$ will then have a proper interpretation, but not~$L_2$. With
  this interpretation, we have that
  \begin{displaymath}
    P(L_3 > v\,|\, L_1 < L_2, L_1 = t_1) = P(\os - \pfs > v\, |\,
    \pfs=\pd < \os) = \exp\left(-\int_\pd^{\pd + v} \alpha_{12}(u;\pd)\dif
      u\right),
  \end{displaymath}
  cf.~Equation~\eqref{eq:condS}. Again, two aspects are apparent from the
  above display. Firstly, the multistate framework is simpler and more
  parsimonious, because it uses only two real life quantities, PFS and OS,
  rather than three quantities~$L_1$, $L_2$ and~$L_3$. Secondly, the
  interpretation of~$L_3$ is unclear, if $L_1 > L_2$. This is in contrast to
  the difference of OS minus PFS, which always has a proper interpretation.

  Neither the difficulties of copula modelling nor those of latent failure
  time modelling are present within the multistate approach. Latent times are
  simply not assumed, such that neither questions of their dependence nor of
  their interpretation arise. All that is modelled are the real life event
  times PFS and OS. Because PFS is modelled as the waiting time in the initial
  state and OS is modelled as the waiting time until absorption, we naturally
  have both PFS~$\le$~OS and equality of PFS and OS with positive probability
  as desired. This also contrasts the multistate approach from using copulas.

\subsection{Transition probabilities and joint distribution of PFS and OS}
\label{sec:jointdist}

For $m, l \in {\cal{S}}$, abbreviate the transition probability $P(X(t) = m |X(s)
= l, \text{past})$, $s\le t$ to be in state $l$ at time $s$ and in state
$m$ at time $t$ by $P_{lm}(s, t; \pd)$, where the argument $\pd$
is only used if the transition probability indeed depends on the time when
progression occurred. For instance, $P_{12}(s, t; \pd)$ conditions on being in
state~$1$ at time $s$ and conditions on the fact that the $0\to 1$ transition
has occurred at time~$\pd$, $\pd \le s$.

Standard considerations (see Aalen et al.\cite{aalen_08}, Appendix A) allow to write the
matrix $(P_{lm}(s,t))_{l,m}$ of transition probabilities as a function of the
transition intensities. In the IDM the upper half of this matrix has non-zero entries and these are
\bean
P_{00}(s, t) &=& \exp\Bl(-\int_s^t \alpha_{01}(u) + \alpha_{02}(u) \dif u\Br), \nonumber \label{eq:P00}\\
P_{11}(s, t; \pd) &=& \exp\Bl(-\int_s^t \alpha_{12}(u; \pd) \dif u\Br), \label{eq:P11} \\
P_{22}(s, t) &=& 1, \label{eq:P22} \nonumber \\
P_{01}(s, t) &=& \int_s^t P_{00}(s, u_{-})\alpha_{01}(u) P_{11}(u, t; u) \dif u, \label{eq:P01} \nonumber \\
P_{12}(s, t; \pd) &=& 1 - P_{11}(s, t; \pd), \label{eq:P12} \nonumber \\
P_{02}(s,
t) 
&=& 1 - (P_{00}(s, t) + P_{01}(s, t) ). \label{eq:P02}
\eean

All other transition probabilities are equal to 0. These transition
probabilities provide a full description of the model in Figure~\ref{fig:msm}
by only assuming existence of the intensities \eqref{eq:csh} and
\eqref{eq:csh12}. The formulae above have interpretations: $P_{00}(s, t)$ and
$P_{11}(s, t; \pd)$ have the form of standard survival functions, and $P(\pfs
> t) = P_{00}(0,t)$. For $P_{01}(s, t)$ one considers the integral of
`infinitesimal probabilities' to move from $0$ to $1$ at time~$u$, $u\in
(s,t]$, --- the term $P_{00}(s, u_{-})\alpha_{01}(u)$ --- and to subsequently
stay in state~$1$ until at least time~$t$ --- the term $P_{11}(u, t; u)$.

The joint distribution PFS and OS in terms of the multistate process is
for $u\le v$
\bean
  P(\pfs \le u, \os \le v) & = & P(X(u) \in \{1,2\}, X(v) =2)\nonumber\\
  {} & = & P(X(u) = 1, X(v) = 2) + P(X(u)=2)\nonumber\\
  {} & = & P(X(v) = 2| X(u) = 1) \cdot P(X(u) = 1|X(0)=0) + P(X(u)=2|X(0)=0) \nonumber\\
  {} & = & P(X(v) = 2| X(u) = 1) \cdot P_{01}(0, u) + P_{02}(0, u). \label{eq:joint}
\eean

If the process is Markov, the above formula simplifies using $P(X(v) =
2| X(u) = 1) = P_{12}(u,v)$, i.e. the expression does not further depend on the time of
progression~$\pd\le u$. For illustration, we provide a closed formula for the joint distribution for $X$ homogeneous Markov in Section~\ref{sec:jointMarkov}. For non-Markov processes,
one has to integrate $P_{12}(u,v;t_1)$ over the conditional distribution of
all possible progression times $\pd\le u$, which makes the final formula
somewhat tedious, see Appendix~\ref{sec:jointnonM}. We reiterate, however,
that $P(X(v) = 2| X(u) = 1)$ can easily be evaluated numerically as explained
in Section~\ref{sub:msm:1}

Many statistics of interest can easily be derived from the joint distribution
of PFS and OS, e.g. Pearson's correlation coefficient as in
Fleischer et al.\cite{fleischer2009statistical} and Li and Zhang\cite{li2015weibull}. 
Precise formulas for these expressions based on the joint and marginal
distribution as well as further dependence measures for bivariate survival
data are discussed in Chapter~4 of Hougaard\cite{hougaard_00} as well as Fan et al.\cite{fan_00}.


Finally, derive the survival function $\sos$ for OS as
 \bean
     \sos(t) \ = \ P(\os > t) &=& P_{00}(0,t) + P_{01}(0,t) \nonumber \\
     &=& \spfs(t) + P_{01}(0,t). \label{eq:Sos}
\eean
Note the independence of $\sos$ of $\pd$, which implies that the generic formula \eqref{eq:Sos} is valid for the broad range of assumptions on $X$ discussed in Section~\ref{sec:assNM}. Closed formulas for the transition probabilities $P_{ij}$, $\spfs$ and $\sos$ for $X(t)$ a semi-Markov, time-inhomogeneous Markov, or homogeneous Markov process and under various parametric assumptions are provided in Section~\ref{sec:closed}.

\subsection{Correlation between PFS and OS}
\label{sub:msm:last}
The joint distribution of PFS and OS together with the corresponding univariate distributions now allows to derive a formula for the correlation between the two random variables PFS and OS. To this end, recall the general correlation formula
\bean
    \Corr(\pfs, \os) &=& \frac{\Cov(\pfs, \os)}{\sqrt{\Var(\pfs) \Var(\os)}} \ = \ \frac{\Ex(\pfs \cdot \os) - \Ex(\pfs) \Ex(\os)}{\sqrt{\Var(\pfs) \Var(\os)}}. \label{eq:corr}
\eean

The formula above can easily be evaluated numerically using the
  simulation approach discussed in Section~\ref{sub:msm:1}. For analytical
  evaluations, mean and variance of $\pfs$ and $\os$ can be derived via the
survival functions \eqref{eq:Spfs} and \eqref{eq:Sos}. One way to compute
$\Ex(\pfs \cdot \os)$ is via the distribution function of the
product. In Appendix~\ref{sec:pfsos}, we show that the survival
function of the random variable $\pfs \cdot \os$ can be written as
\bea
  P(\pfs \cdot \os > t) &=& P(\pfs > \sqrt{t}) + \int_{(0, \sqrt{t}]} P_{11}(u, t / u; u) P(\pfs > u-) \alpha_{01}(u)\,\dif u.
  \label{eq:pfsos}
\eea
Alternatively, the expression \eqref{eq:joint} for the joint distribution
function of PFS and OS can be used together with $\spfs, \sos$, and the
generic formula for the covariance: \bea \Cov(PFS, OS) &=& \int_0^\infty
\int_0^\infty \{P(\pfs > t_1, \os > t_2) - P(\pfs > t_1) P(\os > t_2)\} \dif
t_1 \dif t_2 \eea see e.g. p. 130 in Hougaard\cite{hougaard_00}. 

How to estimate all the above expressions and thus the correlation from given data is discussed in Section~\ref{sec:estcorr}.

Pearson's correlation coefficient is regularly used in the assessment of surrogacy and as discussed in the introduction, methodology for this association measure has been developed in Fleischer et al.\cite{fleischer2009statistical} and Li and Zhang\cite{li2015weibull}. This is why here we also provide a closed formula for this quantity, as well as an application to real data in Section~\ref{sec:estcorr}. We would like to reiterate though that having explicit expressions for the marginal and the joint distribution of PFS and OS allows to derive many dependence measures, as discussed in Section~\ref{sec:jointdist}.



\section{The approaches of Fleischer et al.\ and of Li and Zhang revisited and extended}
\label{sec:fleischli}

We are now fully equipped to revisit the approaches of Fleischer
  et al.~\cite{fleischer2009statistical} and of Li and Zhang
  \cite{li2015weibull}. These authors use latent failure times to generate a
  multistate model for PFS and OS and subsequently model dependence, and
  specifically the correlation, between PFS and OS.
  The assumption of exponentially distributed or time-constant transition
  hazards in Fleischer et al. implies that their model is homogeneous Markov,
  see Section~\ref{sec:assNM}. Furthermore, the assumption of independent
  failure times in Li and Zhang relates to a multistate model which is
  semi-Markov.  Due to these connections of the underlying assumptions of
  these approaches to those in an multistate model, we label those as
  \textit{Homogeneous Markov (Fleischer et al.)} and \textit{Semi-Markov
    Weibull (Li and Zhang)} in Sections~\ref{sec:estcorr} and
  \ref{sec:example}.  In the setup of Section \ref{sec:msm} it is now
  straightforward to drop the hypothetical latent times structure and identify
  the transition hazards following Section~\ref{subsec:latent}. For given
  transition hazards, measures of dependence may easily be evaluated using the
  simulation approach of Section~\ref{sub:msm:1}. Alternatively, one may use
  the formulas in Sections~\ref{sec:jointdist} and~\ref{sub:msm:last} to
  derive closed expressions, but this may turn out to be somewhat tedious, see
  also Appendix~\ref{sec:closed}. For mathematical convenience, Li and Zhang
  \cite{li2015weibull} assume a common shape parameter in their three Weibull
  distributions, but this restriction can also be dropped in our approach.

  Both Fleischer et al.\ (exponential) and Li and Zhang (Weibull) considered
  parametric models, and we provide for parametric inference using maximum
  likelihood methods for counting processes\cite{Kalb:Pren:stat:2002,
    aalen_08} in Section~\ref{sec:inf}. Furthermore, the IDM formulation also
  allows for nonparametric estimation and inference.

  Note that the IDM formulation also allows to some extent to separate the
  assumptions on the underlying process $X$ (time-homogeneity and
  Markovianity) and the assumption about the parametric family for the
  transition intensities. As a matter of fact, assuming exponential transition
  intensities induces $X$ to be homogeneous Markov, but in general, the
  parametric assumption is rather independent of the assumptions on $X$. For
  example, Weibull transition intensities can be assumed for $X$ Markov,
  semi-, or non-Markov. Our chosen IDM approach allows for transparent
  inference based on all the various assumptions on $X$ listed in
  Table~\ref{tab:Ps} and for any (even transition-specific) parametric model
  one is willing to entertain for the transition intensities.

\section{Statistical inference}
\label{sec:inf}

Maximum likelihood estimation in Fleischer et al.\ (exponential) and Li and Zhang (Weibull) relies on the latent failure time structure of the their model together with a parametric assumption. The likelihood is constructed by considering the four potential outcomes for a patient that can occur through censoring and/or events for PFS and OS. In the model entertained by Fleischer et al., the simplicity of the exponential distribution together with the Markov assumption would even allow to derive the parameter estimates via dividing the number of transitions by the waiting time of all individuals for that transition.

Parametric inference in our general multistate model framework outlined in Section~\ref{sec:msm} relies on maximum likelihood estimation for counting processes \cite{Kalb:Pren:stat:2002, aalen_08}. In contrast to the methods used by Fleischer et al.\ (exponential) and Li and Zhang (Weibull), the likelihood is based on the contributions of the three transitions in the IDM.

A general derivation of maximum likelihood estimation for parametric counting
process models is e.g. given in Chapter 5 of Aalen et
al.~\cite{aalen_08}, see also Section VII.6 of~\cite{andersen_93}. In general, these methods hold for a broad class of
stochastic processes. In our case, $X$ is a 3-variate counting process, where
the Markov assumption is only relevant for the $1 \to 2$ transition. As
described in Table~\ref{tab:Ps}, when parametrically modelling the transition
intensities, we can explicitly account for the Markov assumption, implying a
full description of all intensities. This then allows to use the standard
likelihood framework. This approach can easily be generalized to parametric
multistate models that are more general than IDMs, e.g. with more states, and
the (non-)Markov assumption can be taken into account through appropriate
formulation of transition intensities.

  The derivation of the likelihood for an IDM was made explicit for a
  parametric multistate model \cite{andersen_08} and for the example of an
  extended multistate model with exponential transition intensities
  \cite{voncube_17}.

  To derive the likelihood function in the IDM considered here we follow the
  development in Aalen et al. \cite{aalen_08} and assume that $n$ independent
  multistate time-inhomogeneous Markov processes \bea (X_i(t), 0 \le t \le
  C_i; i = 1, \ldots, n) \eea are observed in continuous time. For each $X_i$
  we allow for independent right-censoring at $C_i \le \tau$, with $\tau$ the
  upper time limit of the study, see \cite{aalen_08}, p. 211. The data for
  individual $i$ can then be represented as a multivariate counting process
  $N_{lmi}, l, m \in {\cal{S}}, l < m, t \le C_i$ which counts the number of
  the $i$-th's individual direct transitions from $l$ to $m$ in the interval
  $[0, t]$. The model is then specified, for the assumed parametric transition
  hazards $\alpha_{lm}(t) := \alpha_{lm}(t; \bf{\theta})$ depending on a
  parameter vector $\bf{\theta}$, through the intensity processes
  $\phi_{lmi}(t) = Y_{li}(t) \alpha_{lm}(t; {\bf \theta})$ with $Y_{li}(t) =
  1\{X_i(t-) = l\}$ being an at risk indicator for individual $i$. Denoting by
  $\Delta N_{lmi} = N_{lmi}(t) - N_{lmi}(t-)$ the increment of $N_{lmi}$ at
  time $t$ (i.e. the number of $i$'s transitions from $l$ to $m$ at $t$), the
  likelihood function for estimation of $\bf{\theta}$ can then be written as
  \bea L(\bf{\theta}) &=& \prod_{i=1}^n \prod_{l, m \in {\cal{S}}, l < m}
  \Bl\{ \prod_{0 < t \le \tau} \phi_{lmi}(t)^{\Delta N_{lmi}} \Br\} \exp
  \Bl\{-\int_0^\tau \phi_{lmi}(u) \dif u \Br\}.  \eea In order to obtain a
  maximum likelihood estimate $\bf{\hat \theta}$ of $\bf \theta$ we take the
  logarithm of $L$: \bea \log L(\bf{\theta}) &=& \sum_{i=1}^n \sum_{l, m \in
    {\cal{S}}, l < m} \Bl\{ \sum_{0 < t \le \tau} (\log \alpha_{lm}(t; {\bf
    \theta})) Y_{li}(t) \Delta N_{lmi} \Br\} \exp \Bl\{-\int_0^\tau
  \alpha_{lm}(u; {\bf \theta}) Y_{li}(u)\dif u \Br\} \eea and maximize this
  function over $\bf \theta$. Inference then proceeds via standard maximum
  likelihood theory as outlined in Aalen et al. \cite{aalen_08}, Chapter 5.


Using $\alpha_{lm}(\cdot; \bf{\theta})$ in the format as of Section~\ref{sec:msm}, we get parameter estimates of the transition intensities and probabilities, $S_{PFS}, S_{OS}$, the correlation coefficient $Corr(PFS, OS)$, and the joint distribution of PFS and OS.

Asymptotic covariance matrices can be evaluated based on the delta method, but
we prefer bootstrapping as outlined by Efron\cite{efron_81} or in Bluhmki
  et al.\cite{bluhmki_08}. One reason is that (co)variance formulas become
increasingly tedious, another reason is that the bootstrap is well known to
work well in practice and when compared to asymptotic formulas
\cite{gill_89}. Our preference for bootstrapping is not unlike the idea of
simulation in Section~\ref{sub:msm:1}.

Another advantage of our multistate formulation of the IDM is that it also
allows for non-parametric estimation of all the quantities of interest. {To
  this end, recall that $\alpha_{lm}(t) \dif t$ is the infinitesimal
  conditional transition probability for an individual to be in state $l$ just
  prior to $t$ and transitioning to $m$ in the interval $[t, t + \dif t)$. If
  we observe a transition from $l$ to $m$ for at least one individual $i$ we
  estimate the above probability through the ratio of the sum of increments
  $\sum_{i=1}^n\Delta N_{lmi}(t)$ divided by the total number of individuals
  at risk of the transition out of $l$ just prior to $t$. Summing these
  quantities up yields the Nelson-Aalen estimator of the cumulative transition
  hazards, for $l < m$, in a time-inhomogeneous Markov IDM:
\bea
    \hat A_{lm}(t) &=&  \sum_{s \le t} \frac{\sum_{i=1}^n\Delta N_{lmi}(s)}{\sum_{i=1}^nY_{li}(s)}.
\eea
}
Violations of the Markov property only affect $\hat A_{12}(t)$,
  which would then be estimating an only partly conditional cumulative transition rate,
  i.e., a mixture of the individual transition hazards\cite{Datt:Satt:vali:2001}.
Based on the {corresponding} multivariate Nelson-Aalen
\cite{aalen1978nonparametric} estimator for the matrix of cumulative
transition intensities, the Aalen-Johansen \cite{aalen1978empirical} estimator
of all the transition probabilities in Section~\ref{sec:jointdist} can easily
be derived. {For a Markov IDM, Section 3.4.2 in Aalen et al. \cite{aalen_08}
  provides closed formulas for estimates of these probabilities}.  Plugging in
these estimates in the formulas for $S_{PFS}$, $S_{OS}$, the joint
distribution of PFS and OS, and $Corr(PFS, OS)$ allows for nonparametric
estimation of all these quantities in the IDM framework, either
  in a Markov setting or, using the approach of Putter and
  Spitoni~\cite{putter2016non}, in a non-Markov IDM. However, as is common in
time-to-event outcomes with right-censoring, estimates of survival functions
for PFS and OS generally do not drop down to zero, implying that not the
entire distributions of PFS and OS can be identified nonparametrically. {As a
  consequence, parametric extrapolation would be required to estimate these
  functions and quantities derived from them, specifically the correlation
  coefficient for PFS and OS.}

Despite this limitation, nonparametric estimation provides a valuable tool to check the goodness of fit of an assumed parametric model \cite{aalen_08}.  

Note that the formulas in Section~\ref{sec:jointdist} are also valid if $X$ is non-Markov, so that plugging in estimates of transition probabilities that account for the non-Markovianity yields valid estimates of the above quantities for $X$ non-Markov. Such estimates are e.g. discussed in Meira et al.\cite{meira_06}.

{In our approach, we assume that the date of progression is when it is
  diagnosed. Often, in trials assessing PFS assessments happen at regular
  intervals. How to account for that type of measurements in the likelihood is
  discussed by Zeng et al.\cite{zeng_18}.}

\section{Simulation results for correlation coefficient}
\label{sec:estcorr}
Section~\ref{sub:msm:last} provides formulas that can be used to compute Pearson's correlation coefficient from given data, either through plugging in estimates of $\spfs$ and $\sos$ and integrating numerically (or even analytically in specific situations) or through simulation.

\begin{figure}[h!]
\begin{center}
\setkeys{Gin}{width=0.9\textwidth}
\includegraphics{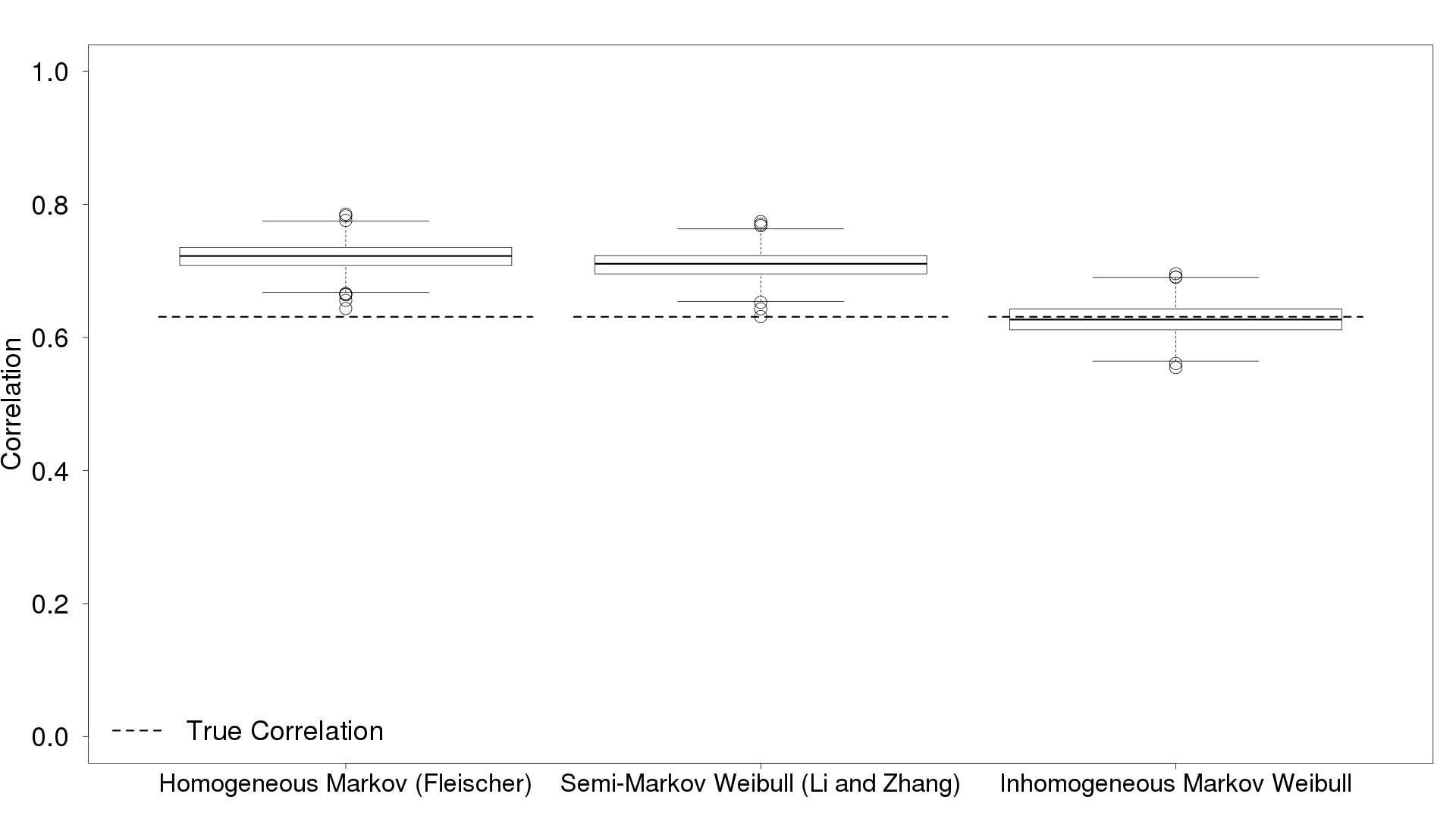}
\end{center}
\caption{Estimated correlation coefficients for Weibull transition intensities (transition-specific shape parameter), time-inhomogeneous Markov process, $1000$ simulation runs. {The horizontal line corresponds to the true correlation, received by plugging in the parametric assumption in the correlation formula in Section~\ref{sub:msm:last}.}}
\label{fig:corr1}
\end{figure}

In Figure~\ref{fig:corr1} we show estimated correlation coefficients from 1000
simulated trials, comparing a homogeneous Markov and Semi-Markov Weibull model
to our proposed approach. For the technical details of the first two we refer
to Fleischer et al.\cite{fleischer2009statistical} and Li and
Zhang\cite{li2015weibull}. In our approach we assume an inhomogeneous Markov
model with Weibull transition hazards, labelled \textit{(Time-) Imhomogeneous
  Markov Weibull} in what follows.


The assumptions for the simulations were as follows: 
For 500 sampled patients, transition intensities were assumed to follow a time-inhomogeneous Markov illness-death model as described in Section~\ref{sub:msm:1}, with Weibull transition intensities as in Table~\ref{tab:weibull}. The parameters were chosen as $\gamma_{01} = 1.5$ and $\lambda_{01} = 0.57$ for the first transition, resulting into a medium hazard of progression which increases over time. With $\gamma_{02} = 0.5$ and $\lambda_{02} = 0.065$, the hazard for death without progression is much lower than the one for progression. As a consequence, on average only about seven percent of patients die without a prior progression. Finally, values of $\gamma_{12} = 0.85$ and $\lambda_{12} = 1.1$ imply a high initial transition hazard from progression to death, but $\alpha_{12}$ decreases over time with the made assumptions, where in this time-inhomogeneous model time is measured from the origin. All this entails that PFS and OS times are generally not too different, meaning that estimated correlation coefficients are relatively high.

Comparing the three estimates of the correlation coefficient in Figure~\ref{fig:corr1}, the models based on the homogeneous
Markov\cite{fleischer2009statistical} and semi-Markov\cite{li2015weibull}
assumption seem not well able to capture that with our choice of the
parameters of $\alpha_{12}$ the hazard to die after progression is lower the
later the progression occurs. As a consequence, both these models are
overestimating the true underlying correlation. 



More results for alternative assumptions on the underlying transition hazards
are provided in a web supplement.

\section{Real data example}
\label{sec:example}


To illustrate the proposed methodology, we applied it to the data of CLEOPATRA\cite{baselga_12}, a large Phase 3 randomized clinical trial in HER2-positive metastatic breast cancer. In the trial, 808 patients were randomized.
 At the primary analysis with a clinical cutoff date of 13th May 2011, the key secondary endpoint investigator-assessed PFS (Inv-PFS) gave an estimated hazard ratio of 0.65 ([0.54, 0.78]). For OS, the hazard ratio was 0.64 ([0.47, 0.88]).
 After an additional interim analysis for OS with clinical cutoff date 14th May 2012, cross-over from the control to the treatment arm was allowed, and about 12\% of the patients initially randomized to the control arm actually crossed over. Using a clinical cutoff of 11th February 2014 for the final analysis on OS, the estimated hazard ratios for Inv-PFS and OS were updated to 0.68 ([0.58, 0.80], based on 604 events) and 0.68 ([0.56, 0.84], 389 deaths), respectively \cite{swain_15}. In what follows, we analyze the data from this last snapshot.

As for the number of transitions, we observe 579 transitions $0 \to 1$, 47 from $0 \to 2$, and 342 from $1 \to 2$.

\subsection{Cumulative hazards}
\label{sub:cumhaz}

\begin{figure}[h!]
\begin{center}
\setkeys{Gin}{width=1\textwidth}
\includegraphics{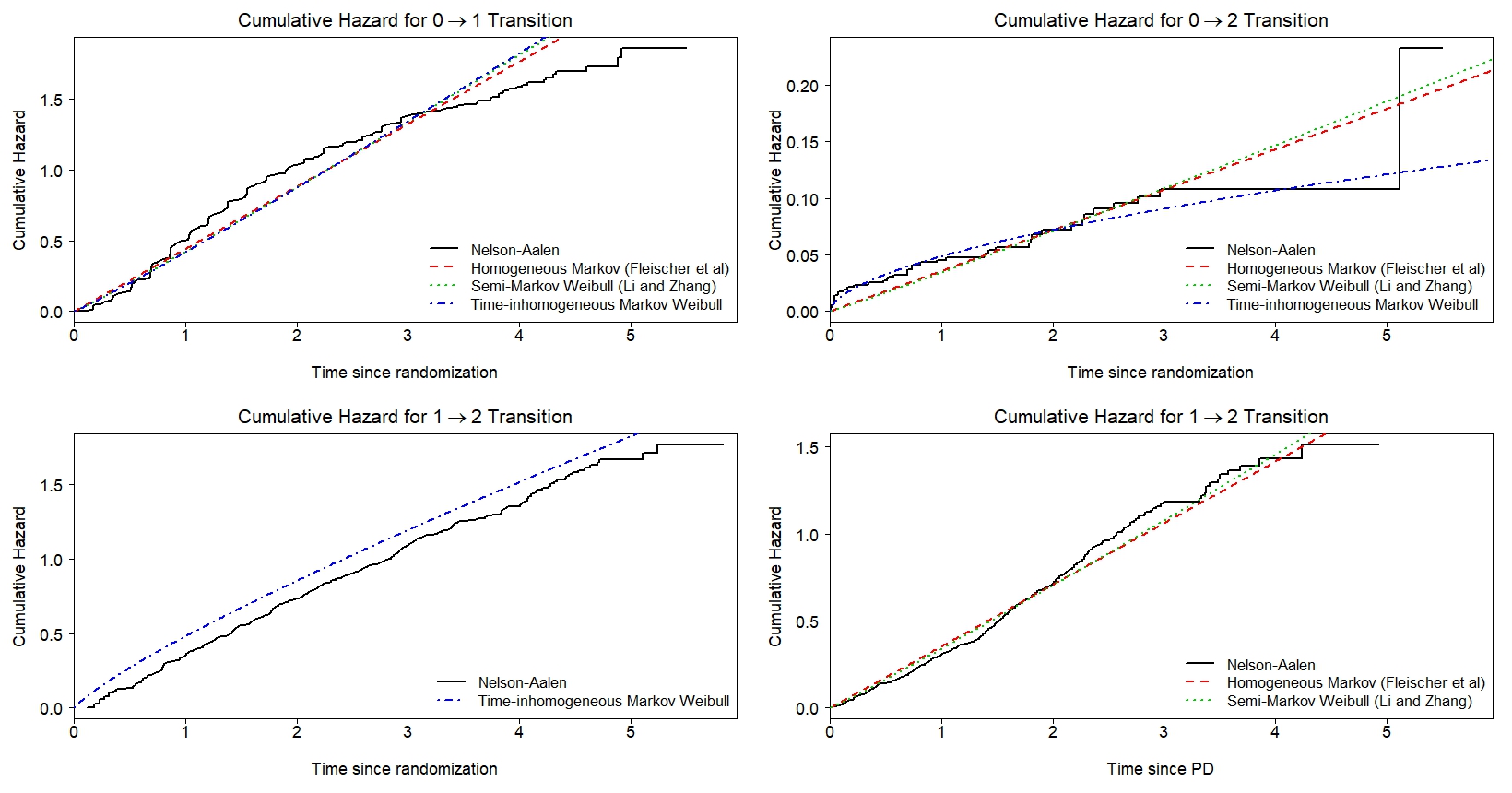}
\end{center}
\caption{CLEOPATRA trial: estimates for cumulative hazards for each transition.}
\label{fig:CleoCumHaz}
\end{figure}

As a first illustration, in Figure~\ref{fig:CleoCumHaz} we provide estimates for the cumulative hazard functions for each transition, under various assumptions, and for all CLEOPATRA patients jointly. The two plots in the upper row and the left-most bottom row plots provide different estimates for the cumulative hazards for each transition in the IDM depicted in Figure~\ref{fig:msm}: the entirely nonparametric Nelson-Aalen estimate, the two homogeneous Markov and semi-Markov parametric estimates assuming an exponential\cite{fleischer2009statistical} or Weibull\cite{li2015weibull} model
and our time-inhomogeneous Markov approach based on the counting process likelihood developed in this paper. For the $0 \to 1$ transition all models provide a very similar and satisfactory fit. For the $0 \to 2$ transition we find that the added flexibility of a model with two more shape parameters provides a better fit, as judged by visually comparing the parametric to the nonparametric estimate. As for the $1 \to2$ transition, we provide two graphs, accounting for the different time-scale of the various models. In the lower-left plot, the time-inhomogeneous Markov model seems to slightly overestimate the cumulative hazard of that transition, likely induced by the shape of the cumulative hazard function of the chosen Weibull distribution, which is an non-decreasing function through the origin. The estimates based on the homogeneous Markov and semi-Markov models by Fleischer et al. and Li and Zhang are depicted in the lower-right plot, with $x$-axis time scale {\it time since progression}.

\subsection{Survival functions}
\label{sub:surv}

\begin{figure}[h!]
\begin{center}
\setkeys{Gin}{width=1\textwidth}
\includegraphics{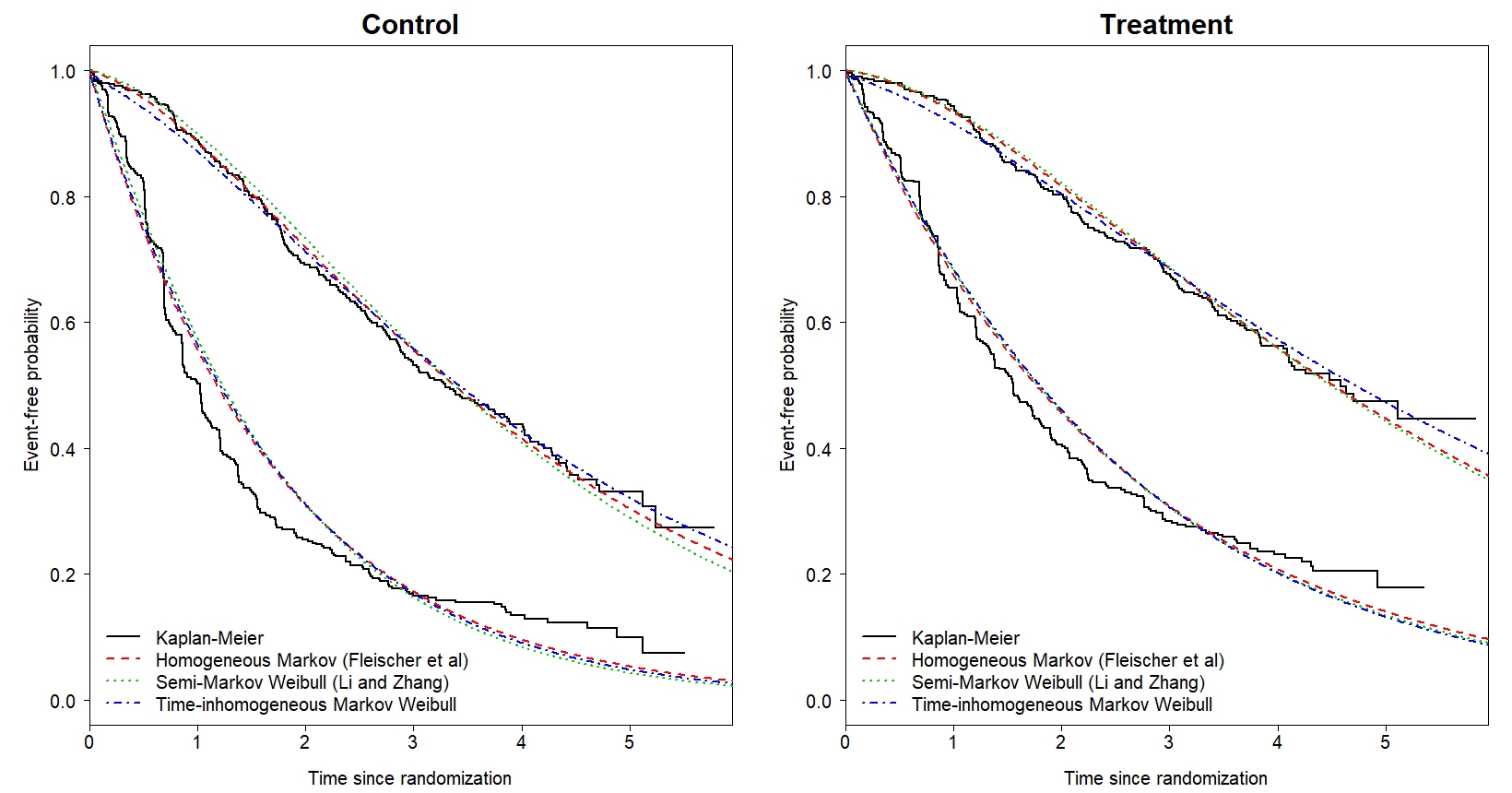}
\end{center}
\caption{CLEOPATRA trial: various estimates of survival functions for OS (upper curves) and PFS (lower curves).}
\label{fig:CleoPFSOSpara}
\end{figure}

Figure~\ref{fig:CleoPFSOSpara} depicts estimates of survival functions for PFS and OS, based on plugging in estimates of cumulative hazards into the formulas for the transition probabilities and plugging these in in turn into the expressions for $\spfs$ and $\sos$. One of the questions that has been explored based on an IDM is to quantify the association between PFS and OS through estimation of Pearson's correlation coefficient\cite{fleischer2009statistical, li2015weibull}. In order to see whether that association is different between control and treatment arm in CLEOPATRA, we split the plot by treatment arms. In general, judging goodness-of-fit of the parametric models through comparison with the nonparametric Kaplan-Meier estimate, we find that generally PFS is very similar for all the methods and in both treatment arms. This is not surprising, for the following reasons: first, the estimates of the cumulative hazard for the $0 \rightarrow 1$ transition are very similar for the different methods. Second, the difference in estimates for the $0 \rightarrow 2$ transition does not relevantly influence $\spfs$ as the number of such transitions is relatively much lower compared to the number of $0 \rightarrow 1$ transitions. Finally, since $\spfs$ in \eqref{eq:Spfs} only depends on the hazards of the $0 \to 1, 2$ transitions it is agnostic to whether we assume a semi-Markov or Markovian model. Thus, similar estimates for $\spfs$ are expected unless the estimated shape parameter of the Weibull model with restricted shape parameter\cite{li2015weibull} for the $0 \rightarrow 1$ transition differs relevantly from the one for our IM model.

The three methods also yield similar estimates for OS, which makes the
  differences seen in the goodness-of-fit in Figure~\ref{fig:CleoCumHaz}
  appear less important. One possible reason for this will be
  that $\mbox{PFS}=\mbox{OS}$ with positive probability.

\subsection{Inference for the correlation coefficient between PFS and OS}
One important aspect when validating PFS as a potential surrogate endpoint for
OS is to analyze their correlation \cite{fleischer2009statistical,
  li2015weibull}. Fleischer et al. discuss that if the correlation is rather
low it might not be reasonable to pursue a joint modelling of PFS and OS but
to consider the two endpoints separately. According to the latter authors,
even more important is the comparison of different correlation values for
different substances, lines of treatment or indications. Examples for
non-small cell lung cancer (NSCLC) \cite{fleischer2009statistical} and
prostate as well as larynx cancer \cite{li2015weibull} have been used to
illustrate the methodology. Estimated correlations between PFS and OS range
from 0.34 for a NSCLC example based on 95 patients in Fleischer et al. to 0.90
in an example based on 356 patients in Li and Zhang.

The estimated correlation coefficients for the CLEOPATRA trial in HER2-positive metastatic breast cancer are provided in Table~\ref{tab:estcorr}. {It is important to note that our proposed methods also allow to provide confidence intervals for these correlation coefficients.} We find that irrespective of the method, the correlation between PFS and OS is lower in the control compared to the treatment arm. A potential reason for this is that after the primary analysis for PFS, crossover from the control to the treatment arm was allowed, potentially confounding the association between PFS and OS in the control arm. As for the various assumptions on $X$ and the parametric transition intensities, estimates for the correlation coefficient are rather comparable and likely not relevantly different taking into account estimation uncertainty.
CLEOPATRA is quite a large trial, with 808 patients, 604 PFS and 389 OS events providing substantial information on PFS and OS. Still, confidence intervals around correlation coefficients have a width of up to 0.15 for the homogeneous Markov and the semi-Markov model (with the latter based on the same shape parameter for all transition intensities) and are even wider for the time-inhomogeneous Markov model with Weibull transition hazards. So, it seems precisely estimating the correlation coefficient is inherently a hard task even for this quite large dataset, and the precision crucially depends on the bias-variance trade-off taken, i.e. whether a more or less restrictive parametric model is entertained. Note that since neither for PFS nor OS the Kaplan-Meier estimate drop down to zero, we are not able to fully nonparametrically estimate the correlation coefficient.

On an absolute scale, the correlation coefficients estimated for CLEOPATRA under different model assumptions are comparable to those Li and Zhang\cite{li2015weibull}.

\renewcommand{\arraystretch}{1.3}
\begin{table}[ht]
\centering
\begin{tabular}{lcc}
                                    & Control arm & Treatment arm \\ \hline
Homogeneous Markov (Fleischer et. al) & 0.543 [0.480; 0.609] & 0.641 [0.571; 0.709]\\ \hline
Semi-Markov Weibull (Li and Zhang) &  0.552 [0.489; 0.615] & 0.644 [0.572; 0.718] \\ \hline
Time-inhomogeneous Markov Weibull & 0.532 [0.434; 0.631] & 0.567 [0.434; 0.680] \\ \hline
\end{tabular}
\caption{CLEOPATRA trial: Estimated correlation coefficients between PFS and OS, including 95\% confidence intervals based on 1000 bootstrap samples.}
\label{tab:estcorr}
\end{table}
\renewcommand{\arraystretch}{1.0}

\section{Discussion}
\label{sec:discussion}
We have suggested multistate modelling to
  jointly model occurrence of PFS and OS. The multistate framework improves on
  the commonly used latent times
  \cite{Buys:Pied:on:1996,goldman_08,fleischer2009statistical,heng_11,crowther2013simulating,li2015weibull,xia_16,nomura_17}
  or copulas
  \cite{burzykowski_01,weir2006statistical,buyse2007progression,fu_13,takeshi2017}
  in that it is the most parsimonious model. It works without hypothetical
  constructions while ensuring the natural properties that $\pfs\le\os$ and
  that equality holds with positive probability. We have demonstrated that the
  joint distribution of PFS and OS can be derived based on a multistate
  model. No further assumptions were required. The multistate model was
  allowed to be non-Markov, and we have only assumed that the model is
  sufficiently smooth in that the transition intensities exist. For
  illustration, we have demonstrated how to express the correlation between
  PFS and OS, which has received some attention recently
  \cite{fleischer2009statistical,li2015weibull}. We have improved on these
  approaches by fitting them
  into our general framework and providing methods of statistical
  inference. The methods were then illustrated using data from a recent
  large randomized clinical trial.

  The real data example also illustrated restrictions of using the correlation
  coefficient to quantify the dependence between PFS and OS. Firstly, and as
  is common in time-to-event outcomes, the Kaplan-Meier curves did not drop
  down to zero for PFS and OS. The implication is that not the entire
  distributions of PFS and OS can be identified nonparametrically, and this is
  in particular true for OS. As a consequence, parametric extrapolation would be required if one wanted to estimate the correlation coefficient based on these nonparametric estimates. Secondly, and almost irrespective of the goodness of the
  parametric fit, the main message from estimating the correlation was that
  the higher PFS, the higher OS. But this is, of course, no surprise, because
  $\pfs\le\os$.

  Our approach allows for any parametric specification of the transition
  intensities, generalizing the use of constant hazards\cite{fleischer2009statistical} or Weibull\cite{li2015weibull} hazards with common shape parameter. If goodness of fit is a
  concern, the transition probabilities of Section~\ref{sec:jointdist} can be
  estimated nonparametrically, either in a time-inhomogeneous Markov model
  \cite{aalen_08} or, provided that censoring is entirely random, even in a
  non-Markov model \cite{putter2016non}. However, this is only possible on the
  time interval of follow-up. If a (partially) nonparametrical analysis is
  desired, on would then need to parametrically model the remainder of the
  multistate distribution, again relying on extrapolation. An alternative
  would be to evaluate the integrals in Section~\ref{sub:msm:last} only up to
  the last observed event time, similar to the concept of restricted mean
  survival, but usefulness of such a restricted dependence measure would need
  to be further investigated. These issues call for further investigating how
  to adequately study the dependence between PFS and OS. We reiterate that
  whatever the choice of such a dependence measure may be, it would be a
  function of the joint distribution of PFS and OS and could hence be studied
  within our framework.

\section{Acknowledgments}
This paper summarizes and extends the results of the first author's MSc thesis
submitted at the University of Ulm. Part of the research for the thesis was
conducted while the first author was still an intern in the Biostatistics Department
of Hoffmann-La Roche in Basel. JB was partially supported by grant
BE 4500/1-2 of the German Research Foundation (DFG).

\appendix

\section{Closed formulas for transition probabilities and for $\spfs$ and
  $\sos$}
\label{sec:closed}

In this section, we provide closed formulas for the transition probabilities $P_{ij}$ and survival functions for PFS and OS for $X(t)$ a semi-Markov, time-inhomogeneous Markov, and homogeneous Markov process. The non-Markov case requires further assumptions on how transition intensities depend on $\pd$, and we will not pursue this further here.

As discussed in Section~\ref{sec:msm}, in the case of $X(t)$ being homogeneous Markov then the transition intensities are constant in time since origin (and independent of $\pd$) and thus exponential. Reversely put, if we assume ``homogeneous'' (i.e. $\alpha_{ij}$ independent of $t$ and $\pd$) exponentially distributed transition intensities, then the homogeneous and time-inhomogeneous Markov as well as the semi-Markov model all coincide. If we allow $\alpha_{12}$ to depend on $\pd$ then a more general non-Markov, or Markov extension in the terminology of Hougaard\cite{hougaard_00}, assumption for $X(t)$ has to be entertained.

Fleischer et al.\ \cite{fleischer2009statistical} (exponential)
and Li and Zhang \cite{li2015weibull} (Weibull) assumed
parametric models in a latent failure time model for the modelling of PFS and
OS. As much as possible at all, we compare our results from a multistate model
for $\spfs$ and $\sos$ to what they received.

Table~\ref{tab:Ps} provides general formulas for transition probabilities under the various assumptions on $X(t)$.

\begin{landscape}
\begin{table}[h]
\renewcommand{\arraystretch}{2}
\begin{center}
\begin{tabular}{c|c|c|c|c}
         & Non-Markov & Semi-Markov & Time-inhomogeneous & Homogeneous \\[-11pt]
         &            &                           & Markov             & Markov (exponential) \\ \hline
$P_{00}$ & \multicolumn{3}{c|}{$\exp\Bl(-\int_s^t \alpha_{01}(u) + \alpha_{02}(u) \dif u\Br)$} & $P_{00}^{\text{TH}}(s,t)$ below \\ \hline
$P_{11}$ & $P_{11}^{\text{NM}}(s, t;\pd) := \exp\Bl(-\int_s^t \alpha_{12}(u; t_1) \dif u\Br)$ & $P_{11}^{\text{SM}}(s, t;\pd) := \exp\Bl(-\int_s^t \alpha_{12}(u - t_1) \dif u\Br)$ & $P_{11}^{\text{TI}}(s, t) := \exp\Bl(-\int_s^t \alpha_{12}(u) \dif u\Br)$ & $P_{11}^{\text{TH}}(s, t)$ below \\ \hline
$P_{22}$ & \multicolumn{4}{c}{1}                                             \\ \hline
$P_{01}$ & $\int_s^t P_{00}(s, u_{-})\alpha_{01}(u) P_{11}^\text{NM}(u, t; u) \dif u$ & $\int_s^t P_{00}(s, u_{-})\alpha_{01}(u) P_{11}^\text{SM}(u, t; u) \dif u$ & $\int_s^t P_{00}(s, u_{-})\alpha_{01}(u) P_{11}^\text{TI}(u, t) \dif u$ & $P_{01}^{\text{TH}}(s, t)$ below \\ \hline
$P_{12}$ & $1-P_{11}^{\text{NM}}(s, t;\pd)$ & $1-P_{11}^{\text{SM}}(s, t;\pd)$ & $1-P_{11}^{\text{TI}}(s, t)$ & $1-P_{11}^{\text{TH}}(s, t)$ \\ \hline
$P_{02}$ & \multicolumn{4}{c}{combine \eqref{eq:P02} with the column-wise expressions for $P_{00}$ and $P_{11}$} \\ \hline
\end{tabular}
\caption{Formulas for transition probabilities $P_{lm}(s,t)$, for various assumptions on the process $X(t)$. See Appendix~\ref{sub:expundhop} below for the homogeneous Markov case.}
\label{tab:Ps}
\renewcommand{\arraystretch}{1.0}
\end{center}
\end{table}
\end{landscape}

\subsection{Exponential (time-constant) transition hazards}
\label{sub:expundhop}

As for the homogeneous Markov or exponential case, we assume that the transition hazards in our multistate model in Figure~\ref{fig:msm} are constant and independent of $\pd$:
 \bea
	\alpha_{01}(t) \ = \ \lambda_{01}, \hspace{1cm} 	\alpha_{02}(t) \ = \ \lambda_{02}, \hspace{1cm} \alpha_{12}(t; \pd)= \lambda_{12}.
\eea
Then, abbreviating $\lambda_{012} := \lambda_{12} - \lambda_{01} - \lambda_{02}$,
\bea
    P_{00}^{\text{TH}}(s, t) &=& \exp\Bl(-\int_s^t \lambda_{01} + \lambda_{02} \dif u\Br) \ = \ \exp\Bl(-(\lambda_{01} + \lambda_{02})(t-s)\Br), \\
    P_{11}^{\text{TH}}(s, t) &=& \exp\Bl(-\int_s^t \lambda_{12} \dif u\Br) \ = \ \exp\Bl( -\lambda_{12} (t-s)\Br), \\
    P_{01}^{\text{TH}}(s, t) &=& \int_s^t P_{00}^{\text{TH}}(s, u_{-})\lambda_{01} P_{11}^\text{TH}(u, t) \dif u \ = \ \lambda_{01} \int_s^t \exp\Bl(-(\lambda_{01} + \lambda_{02})(u_{-}-s)\Br) \exp\Bl( -\lambda_{12} (t-u)\Br) \dif u \\
    &=& \lambda_{01} \exp\Bl((\lambda_{01} + \lambda_{02})s -\lambda_{12} t\Br) \int_s^t \exp(\lambda_{012}u)  \dif u \\
    &=& \lambda_{01} \lambda_{012}^{-1} \Bl[\exp\Bl(-(\lambda_{01} + \lambda_{02})(t-s)\Br) - \exp\Bl(-\lambda_{12}(t- s)\Br)\Br].
\eea We note that all these quantities depend on the interval limits $s$ and $t$ only through the difference $t-s$, as is expected for a time-homogeneous Markov model.

Plugging the expressions from Table~\ref{tab:Ps} in \eqref{eq:Spfs} and \eqref{eq:Sos} we get
\bea
	S_{PFS}^{\text{TH}}(t) &=& P_{00}^{\text{TH}}(0, t) \ = \ \exp \Bl( - (\lambda_{01}+\lambda_{02}) t\Br)
\eea and
\bea
	S_{OS}^{\text{TH}}(t) \ = \ S_{PFS}^{\text{TH}}(t) + P_{01}^{\text{TH}}(0, t)
&=& \frac{\lambda_{12}-\lambda_{02}}{\lambda_{012}} \exp\Bl(-(\lambda_{01}+\lambda_{02})t\Br) + \frac{\lambda_{01}}{\lambda_{012}}\exp(-\alpha_{12}t).
\eea
Comparing $S_{PFS}^{\text{TH}}$ and $S_{OS}^{\text{TH}}$ to those in Theorem~5 in Fleischer et al.\cite{fleischer2009statistical} we find that these formulas are identical to those derived using the latent failure time assumption.

\subsection{Weibull transition hazards}
\label{sub:weibhaz}

If a Weibull model should be entertained for the transition hazards, then we can either specify a time-inhomogeneous Markov or semi-Markov model as in Table~\ref{tab:weibull}.

\begin{table}[h]
\renewcommand{\arraystretch}{1.5}
\begin{center}
\begin{tabular}{c|c|c|c}
                         & $\alpha_{01}$ & $\alpha_{02}$ & $\alpha_{12}$ \\ \hline
time-inhomogeneous Markov & \multirow{2}{*}{$\alpha_{01}(t) = \lambda_{01} \cdot \gamma_{01} t^{\gamma_{01}-1}$}  &
\multirow{2}{*}{$\alpha_{02}(t) = \lambda_{02} \cdot \gamma_{02} t^{\gamma_{02}-1}$} & $\alpha_{12}(t) = \lambda_{12} \cdot \gamma_{12} t^{\gamma_{12}-1}$ \\ \cline{4-4}
semi-Markov   &               &               & $\alpha_{12}(t; \pd) = \lambda_{12} \cdot \gamma_{12} (t-\pd)^{\gamma_{12}-1}$ \\ \hline
\end{tabular}
\caption{Weibull transition intensities.}
\label{tab:weibull}
\renewcommand{\arraystretch}{1.0}
\end{center}
\end{table}

Note that using the multistate formulation, Weibull intensities with transition-specific shape parameters can be used, no restriction to a common shape parameter as in Li and Zhang \cite{li2015weibull} is needed to be able to derive closed formulas.

For the time-inhomogeneous model, plugging in into \eqref{eq:Spfs} and \eqref{eq:Sos} yields, after some straightforward algebraic manipulations,
\bea
	S_{PFS}^{\text{TI}}(t) \ = \ P_{00}^{\text{SM}}(0, t) &=& \exp ( - \lambda_{01} t^{\gamma_{01}} - \lambda_{02} t^{\gamma_{02}})
\eea and
\bea
	S_{OS}^{\text{TI}}(t) &=& S_{PFS}^{\text{TI}}(t) + P_{01}^{\text{TI}}(0, t) \\
                          &=& S_{PFS}^{\text{TI}}(t) + \lambda_{01} \gamma_{01} \int_0^t u^{\gamma_{01} - 1} \cdot \exp \Bl(-\lambda_{01}u^{\gamma_{01}}-\lambda_{02}u^{\gamma_{02}} - \lambda_{12}  (t^{\gamma_{12}}-u^{\gamma_{12}})\Br) \dif u
\eea for the survival function for OS. As for the semi-Markov model, the hazard $\alpha_{12} = \alpha_{12}(t; \pd)$ for the transition from PD to death not only depends on the actual time $t$, but also on the time $\pd$ of the transition out of the initial state. Similar manipulations yield $S_{PFS}^{\text{SM}} \ = \ S_{PFS}^{\text{TI}}$ for the survival function for PFS and
\bea
	S_{OS}^{\text{SM}}(t) &=& S_{PFS}^{\text{SM}}(t) + P_{01}^{\text{SM}}(0, t) \\
                          &=& S_{PFS}^{\text{SM}}(t) + \lambda_{01} \gamma_{01} \int_0^t u^{\gamma_{01} - 1} \cdot \exp \Bl(-\lambda_{01}u^{\gamma_{01}}-\lambda_{02}u^{\gamma_{02}} - \lambda_{12}  (t-u)^{\gamma_{12}}\Br) \dif u
\eea for the survival function for OS. Now, how do these expressions compare to those derived under a latent failure time assumption in Li and Zhang\cite{li2015weibull} when setting $\gamma := \gamma_{01} = \gamma_{02} = \gamma_{12}$? The survival function $S_{PFS}^{\text{TI}} = S_{PFS}^{\text{SM}}$ for PFS simplifies to that in the latter paper as does $S_{OS}^{\text{SM}}$.
We would like to emphasize that the fact that the latent failure time and multistate model formulation lead to the same expressions for the semi-Markov case does {\it not} imply that the assumptions on $X(t)$ are identical for the latent failure time and the multistate model.

To conclude this section we note that \eqref{eq:Spfs} and \eqref{eq:Sos} are general formulas based on the IDM. Closed formulas for other parametric models can easily be derived from these.

\section{Joint distribution of PFS and OS if $X$ is Markov}
\label{sec:jointMarkov}
Equation \eqref{eq:joint} provides the expression of the joint distribution of PFS and OS for a general, not necessarily Markov, process $X$. If we assume $X$ to be Markov, then $P_{12}(u, v; \pd)$ simplifies to $P_{12}(u, v)$ so that we can write (\ref{eq:joint}) as
\bea
 P(\pfs \le u, \os \le v) & = & 1-P_{00}(0, u) -P_{01}(0, u)P_{11}(u,v).
\eea More specifically, if $X$ is homogeneous Markov we can plug in the formulas derived in Section~\ref{sub:expundhop} to get
\bea
  P(\pfs \le u, \os \le v) & = & 1 - \lambda_{01} \lambda_{012}^{-1} \exp(-\lambda_{12}v) + \exp(-\lambda_{012}u)\Bl(\lambda_{01} \lambda_{012}^{-1}\exp(-\lambda_{12}v) - \exp(-\lambda_{12}u)\Br).
\eea
Corresponding formulas for alternative assumptions on $X$ can be derived analogously.

\section{Joint distribution of PFS and OS if $X$ is non-Markov}\label{sec:jointnonM}
Here, we provide the general expression for the joint distribution of PFS and OS if $X$ is non-Markov. To this end, we need to evaluate $P(X(v) = 2| X(u) = 1)$ under that assumption. The following computations are repeatedly using the formula of total probability and conditioning. Assuming $u \le v$ we have
\bean
    P(X(v) = 2| X(u) = 1) &=& \int_0^u P(X(v) = 2 | X(u) = 1, \pfs = t_1) \dif P(\pfs \le t_1 | X(u) = 1) \nonumber \\
    &=& \int_0^u P_{12}(u, v ; t_1) \dif P(\pfs \le t_1 | X(u) = 1). \label{eq:P12nm}
\eean
In~\ref{eq:P12nm}, integration is w.r.t.\ $t_1$ and over the time
interval~$(0,u]$. Now, for $t_1 \le u$, we have that
\bean
    P(\pfs \le t_1 | X(u) = 1) &=& P(X(t_1) \in \{1, 2\} | X(u)=1) \nonumber \\
    &=& P(X(t_1) = 1 | X(u) = 1) \label{eq:t1u} \\
    &=& P(X(t_1) = 1, X(u) = 1) / P(X(u) = 1) \nonumber \\
    &=& P(X(t_1) = 1, X(u) = 1) / P_{01}(0, u), \label{eq:t1u2}
\eean where \eqref{eq:t1u} follows, because given the process is
  still in the intermediate state at~$u$, $u\ge t_1$, $X(t_1)=2$ is impossible.
As for the nominator, 
\bea
    P(X(t_1) = 1, X(u) = 1) &=& \int_0^{t_1} P(X(t_1) = 1, X(u) = 1 | \pfs = s) \dif P(\pfs \le s) \\
    &=& \int_0^{t_1} P(X(u) = 1 | X(t_1) = 1 , \pfs = s) P(X(t_1) = 1 | \pfs = s) \dif P(\pfs \le s) \\
    &=& \int_0^{t_1} P_{11}(t_1, u; s) P(X(t_1) = 1 | \pfs = s) \dif P(\pfs \le s) \ \ \ \text{by \eqref{eq:P11}.}\\
\eea It remains to derive $P(X(t_1) = 1 | \pfs = s)$ for $s\in
  (0, t_1]$. To this end,
\bea
    P_{11}(s, t_1;s) &=& P(X(t_1) = 1 | X(s) = 1, \pfs = s) \\
    &=&P(X(t_1) = 1, X(s) = 1 | \pfs = s) / P(X(s) = 1 | \pfs = s) \\
    &=&P(X(t_1) = 1 | \pfs = s) / P(X(s) = 1 | \pfs = s)
\eea because the event $\{X(s) = 1\}$ is implied by the condition $\{\pfs =
s\}$ together with noting that $X$ is still in state 1 at $t_1$,
  $t_1 \ge s$. Now, the denominator in the last display is
  \begin{displaymath}
    P(X(s) = 1 | \pfs = s) = P(X(\pfs) = 1 | \pfs = s) =
    \frac{\alpha_{01}(s)}{\alpha_{01}(s) + \alpha_{02}(s)},
  \end{displaymath}
  see our discussion on how to simulate PFS-OS trajectories in Section~\ref{sub:msm:1}. Rearranging gives
\bea
    P(X(t_1) = 1 | \pfs = s) &=& \frac{\alpha_{01}(s)}{\alpha_{01}(s) + \alpha_{02}(s)} P_{11}(s, t_1;s).
\eea The last display has an interpretation: Conditional on PFS
  at time~$s$ one is in the progression state at a later time~$t_1$, if,
  firstly, one actually moves into the progression state at the time of
  PFS. Given a transition into the progression state at this time, secondly,
  one has to remain in the intermediate state of the model until time~$t_1$. Putting everything together, we thus have
\bea
   P(\pfs \le t_1 | X(u) = 1) &=& P_{01}(0, u)^{-1} \int_0^{t_1} P_{11}(t_1, u; s) P_{11}(s, t_1;s) \frac{\alpha_{01}(s)}{\alpha_{01}(s) + \alpha_{02}(s)} \dif P(\pfs \le s).
\eea Plugging this distribution function into \eqref{eq:P12nm} finally allows
to compute $P(X(v) = 2| X(u) = 1)$ under a non-Markov
assumption. We reiterate that our simulation approach in
  Section~\ref{sub:msm:1} may be a convenient alternative to evaluating these
  somewhat tedious formulas.

\section{Distribution function of $\pfs \cdot \os$}
\label{sec:pfsos}
We have that
  \bea
    P(\pfs \cdot \os > t) &=& P(\pfs > \sqrt{t}) + P(\pfs \le \sqrt{t}, \pfs \cdot \os > t).
  \eea
  When $\pfs \le \sqrt{t}, \pfs \cdot \os > t$ is only possible, if
  $X(\pfs)=1$. Hence, we can express the last probability as
  \bea
    P(\pfs \le \sqrt{t}, \pfs \cdot \os > t) & = & \int P(\pfs \le \sqrt{t}, \pfs \cdot \os > t\, |\, \pfs = u, X(\pfs)=1)\,\dif P(\pfs\le u, X(\pfs)=1)\\
    {} & = & \int_{(0, \sqrt{t}]} P(\os > t / u\, |\, \pfs = u, X(\pfs)=1) \,\dif P(\pfs\le u, X(\pfs)=1)\\
    {} & = & \int_{(0, \sqrt{t}]} P_{11}(u, t/u; u) \,\dif P(\pfs\le u, X(\pfs)=1).
  \eea
 It remains to compute $P(\pfs \le u, X(\pfs) = 1)$, but this is nothing else than the cumulative incidence
function of a competing risk, i.e. the expected proportion of individuals experiencing progression over the course of time. It
can be computed according to
\bea
    P(\pfs \le u, X(\pfs) = 1) &=& \int_0^u P(\pfs > s-) \alpha_{0{\color{blue}1}}(s) \dif s \nonumber \\
    &=& \int_0^u \spfs(s-) \alpha_{0{\color{blue}1}}(s) \dif s, \label{eq:cuminc}
\eea see e.g. Equation (3.11) in Beyersmann et al.\cite{bam}.

\bibliography{literatur}

\begin{thebibliography}{10}

\bibitem{fda_endpoints}
{U.S. Food and Drug Administration} . {Guidance for Industry: Clinical Trial
  Endpoints for the Approval of Cancer Drugs and Biologics.}2007.

\bibitem{pazdur_08}
Pazdur R.. {{E}ndpoints for assessing drug activity in clinical trials}.  {\it
  Oncologist. }2008;13 Suppl 2:19--21.

\bibitem{george_16}
George S.~L., Wang X., Pang H.. {\it Cancer Clinical Trials: Current and
  Controversial Issues in Design and Analysis}.
\newblock CRC Biostatistics SeriesChapman \& Hall; 2016.

\bibitem{saad_16}
Saad E.~D., Buyse M.. {{S}tatistical controversies in clinical research: end
  points other than overall survival are vital for regulatory approval of
  anticancer agents}.  {\it Ann. Oncol.. }2016;27(3):373--378.

\bibitem{burzykowski_01}
Burzykowski Tomasz, Molenberghs Geert, Buyse Marc, Geys Helena, Renard Didier.
  Validation of surrogate end points in multiple randomized clinical trials
  with failure time end points.  {\it Journal of the Royal Statistical Society:
  Series C (Applied Statistics). }2001;50(4):405--422.

\bibitem{buyse_16}
Buyse M., Molenberghs G., Paoletti X., et al. {{S}tatistical evaluation of
  surrogate endpoints with examples from cancer clinical trials}.  {\it Biom J.
  }2016;58(1):104--132.

\bibitem{prasad2015}
V~Prasad, C~Kim, M~Burotto, A~Vandross. The strength of association between
  surrogate end points and survival in oncology: A systematic review of
  trial-level meta-analyses.  {\it JAMA Internal Medicine.
  }2015;175(8):1389--1398.

\bibitem{fleischer2009statistical}
Fleischer Frank, Gaschler-Markefski Birgit, Bluhmki Erich. A statistical model
  for the dependence between progression-free survival and overall survival.
  {\it Statistics in Medicine. }2009;28(21):2669--2686.

\bibitem{li2015weibull}
Li~Yimei, Zhang Qiang. A Weibull multi-state model for the dependence of
  progression-free survival and overall survival.  {\it Statistics in Medicine.
  }2015;34(17):2497--2513.

\bibitem{fu_13}
Fu~H., Wang Y., Liu J., Kulkarni P.~M., Melemed A.~S.. {{J}oint modeling of
  progression-free survival and overall survival by a {B}ayesian normal induced
  copula estimation model}.  {\it Stat Med. }2013;32(2):240--254.

\bibitem{hougaard_00}
Hougaard P.. {\it Analysis of Multivariate Survival Data}.
\newblock Statistics for Biology and HealthSpringer New York; 2000.

\bibitem{aalen_08}
Aalen Odd, Borgan Ornulf, Gjessing Haakon. {\it Survival and event history
  analysis: a process point of view}.
\newblock Springer Science \& Business Media; 2008.

\bibitem{weber_18}
Weber Enya~M., Titman Andrew~C.. Quantifying the association between
  progression-free survival and overall survival in oncology trials using
  Kendall's $\tau$.  {\it Statistics in Medicine. };0(0).

\bibitem{Ande:Keid:mult:2002}
Andersen PK, Keiding N. Multi-state models for event history analysis.  {\it
  Statistical Methods in Medical Research. }2002;11(2):91--115.

\bibitem{bam}
Beyersmann Jan, Allignol Arthur, Schumacher Martin. {\it Competing Risks and
  Multistate Models with R}.
\newblock Springer, New York; 2012.

\bibitem{beyersmann08:_simul}
Beyersmann J, Latouche A, Buchholz A, Schumacher M. Simulating competing risks
  data in survival analysis..  {\it Statistics in Medicine. }2009;28:956--971.

\bibitem{putter_07}
Putter H., Fiocco M., Geskus R.~B.. {{T}utorial in biostatistics: competing
  risks and multi-state models}.  {\it Stat Med. }2007;26(11):2389--2430.

\bibitem{andersen_08}
Andersen P.~K., Pohar~Perme M.. {{I}nference for outcome probabilities in
  multi-state models}.  {\it Lifetime Data Anal. }2008;14(4):405--431.

\bibitem{andersen_17}
Andersen Per~Kragh. Life years lost among patients with a given disease.  {\it
  Statistics in Medicine. }2017;36(22):3573--3582.

\bibitem{Buys:Pied:on:1996}
Buyse M., Piedbois P.. On the relationship between response to treatment and
  survival time.  {\it Statistics in Medicine. }1996;15:2797--2812.

\bibitem{goldman_08}
Goldman B., LeBlanc M., Crowley J.. {{I}nterim futility analysis with
  intermediate endpoints}.  {\it Clin Trials. }2008;5(1):14--22.

\bibitem{heng_11}
Heng D.~Y., Xie W., Bjarnason G.~A., et al. {{P}rogression-free survival as a
  predictor of overall survival in metastatic renal cell carcinoma treated with
  contemporary targeted therapy}.  {\it Cancer. }2011;117(12):2637--2642.

\bibitem{crowther2013simulating}
Crowther Michael~J, Lambert Paul~C. Simulating biologically plausible complex
  survival data.  {\it Statistics in Medicine. }2013;32(23):4118--4134.

\bibitem{xia_16}
Xia F., George S.~L., Wang X.. {{A} {M}ulti-state {M}odel for {D}esigning
  {C}linical {T}rials for {T}esting {O}verall {S}urvival {A}llowing for
  {C}rossover after {P}rogression}.  {\it Stat Biopharm Res. }2016;8(1):12--21.

\bibitem{nomura_17}
Nomura S., Hirakawa A., Hamada C.. {{S}ample size determination for the current
  strategy in oncology phase 3 trials that tests progression-free survival and
  overall survival in a two-stage design framework}.  {\it J Biopharm Stat.
  }2017;.

\bibitem{weir2006statistical}
Weir Christopher~J, Walley Rosalind~J. Statistical evaluation of biomarkers as
  surrogate endpoints: a literature review.  {\it Statistics in medicine.
  }2006;25(2):183--203.

\bibitem{buyse2007progression}
Buyse Marc, Burzykowski Tomasz, Carroll Kevin, et al. Progression-free survival
  is a surrogate for survival in advanced colorectal cancer.  {\it Journal of
  clinical oncology. }2007;25(33):5218--5224.

\bibitem{takeshi2017}
Emura T., Nakatochi M., Murotani K., Rondeau V.. {{A} joint frailty-copula
  model between tumour progression and death for meta-analysis}.  {\it Stat
  Methods Med Res. }2017;26(6):2649--2666.

\bibitem{Kalb:Pren:stat:2002}
Kalbfleisch J., Prentice Ross. {\it The Statistical Analysis of Failure Time
  Data. 2nd Ed.}
\newblock Wiley, Hoboken; 2002.

\bibitem{Aale:dyna:1987}
Aalen Odd. {Dynamic modelling and causality}.  {\it Scandinavian Actuarial
  Journal. }1987;:177--190.

\bibitem{chiang1991competing}
Chiang Chin~Long. Competing risks in mortality analysis.  {\it Annual review of
  public health. }1991;12(1):281--307.

\bibitem{fan_00}
Fan J., Prentice R.~L., Hsu L.. A class of weighted dependence measures for
  bivariate failure time data.  {\it Journal of the Royal Statistical Society:
  Series B (Statistical Methodology). }2000;62(1):181--190.

\bibitem{andersen_93}
Andersen Per~Kragh, Borgan Ornulf, Gill Richard~D., Keiding Niels. {\it
  Statistical Models Based on Counting Processes}.
\newblock Springer; 1993.

\bibitem{voncube_17}
Cube M., Schumacher M., Wolkewitz M.. {{B}asic parametric analysis for a
  multi-state model in hospital epidemiology}.  {\it BMC Med Res Methodol.
  }2017;17(1):111.

\bibitem{efron_81}
Efron B.. {{C}ensored {D}ata and the {B}ootstrap}.  {\it J. Amer. Statist.
  Assoc.. }1981;76:312--319.

\bibitem{bluhmki_08}
Bluhmki Tobias, Schmoor Claudia, Dobler Dennis, et al. A wild bootstrap
  approach for the Aalen-Johansen estimator.  {\it Biometrics. };0(0).

\bibitem{gill_89}
Gill Richard~D., Wellner Jon~A., Praestgaard Jens. Non- and Semi-Parametric
  Maximum Likelihood Estimators and the Von Mises Method (Part 1) [with
  Discussion and Reply].  {\it Scandinavian Journal of Statistics.
  }1989;16(2):97-128.

\bibitem{Datt:Satt:vali:2001}
Datta Somnath, Satten Glen~A.. Validity of the {A}alen-{J}ohansen estimators of
  stage occupation probabilities and {N}elson-{A}alen estimators of integrated
  transition hazards for non-{M}arkov models.  {\it Statistics and Probability
  Letters. }2001;55(4):403--411.

\bibitem{aalen1978nonparametric}
Aalen Odd. Nonparametric inference for a family of counting processes.  {\it
  The Annals of Statistics. }1978;:701--726.

\bibitem{aalen1978empirical}
Aalen Odd~O, Johansen S{\o}ren. An empirical transition matrix for
  non-homogeneous Markov chains based on censored observations.  {\it
  Scandinavian Journal of Statistics. }1978;:141--150.

\bibitem{putter2016non}
Putter H., Spitoni C.. {{N}on-parametric estimation of transition probabilities
  in non-{M}arkov multi-state models: {T}he landmark {A}alen-{J}ohansen
  estimator}.  {\it Stat Methods Med Res. }2018;27(7):2081--2092.

\bibitem{meira_06}
Meira-Machado Lu{\'i}s, U{\~{n}}a-{\'A}lvarez Jacobo, Cadarso-Su{\'a}rez
  Carmen. Nonparametric estimation of transition probabilities in a non-Markov
  illness--death model.  {\it Lifetime Data Analysis. }2006;12(3):325--344.

\bibitem{zeng_18}
Zeng Leilei, Cook Richard~J., Lee Ker-Ai. Design of cancer trials based on
  progression-free survival with intermittent assessment.  {\it Statistics in
  Medicine. };37(12):1947-1959.

\bibitem{baselga_12}
Baselga J., Cortes J., Kim S.~B., et al. {{P}ertuzumab plus trastuzumab plus
  docetaxel for metastatic breast cancer}.  {\it N. Engl. J. Med..
  }2012;366(2):109--119.

\bibitem{swain_15}
Swain S.~M., Baselga J., Kim S.~B., et al. {{P}ertuzumab, trastuzumab, and
  docetaxel in {H}{E}{R}2-positive metastatic breast cancer}.  {\it N. Engl. J.
  Med.. }2015;372(8):724--734.

\end{thebibliography}


\begin{thebibliography}{1}

\bibitem{meller_18}
Meller M., Beyersmann J., Rufibach K.. {\it Joint modelling of progression-free
  and overall survival and computation of correlation measures. }: Department
  of Biostatistics, F. Hoffmann-La Roche, Basel; 2018.

\bibitem{fleischer2009statistical}
Fleischer Frank, Gaschler-Markefski Birgit, Bluhmki Erich. A statistical model
  for the dependence between progression-free survival and overall survival.
  {\it Statistics in Medicine. }2009;28(21):2669--2686.

\bibitem{li2015weibull}
Li~Yimei, Zhang Qiang. A Weibull multi-state model for the dependence of
  progression-free survival and overall survival.  {\it Statistics in Medicine.
  }2015;34(17):2497--2513.

\bibitem{Ande:Keid:mult:2002}
Andersen PK, Keiding N. Multi-state models for event history analysis.  {\it
  Statistical Methods in Medical Research. }2002;11(2):91--115.

\bibitem{baselga_12}
Baselga J., Cortes J., Kim S.~B., et al. {{P}ertuzumab plus trastuzumab plus
  docetaxel for metastatic breast cancer}.  {\it N. Engl. J. Med..
  }2012;366(2):109--119.

\end{thebibliography}


\clearpage

\end{document}


\title{Web-based Supporting Materials for ``Joint modelling of progression-free and overall survival and
  computation of correlation measures''}

\author[1]{Matthias Meller*}

\author[2]{Jan Beyersmann}

\author[3]{Kaspar Rufibach}

\authormark{Meller \textsc{et al}}

\address[1, 3]{\orgdiv{Department of Biostatistics}, \orgname{F. Hoffmann-La Roche Ltd},
  \orgaddress{\state{Basel}, \country{Switzerland}}}

\address[2]{\orgdiv{Institute of Statistics}, \orgname{Ulm University}, \orgaddress{\state{Helmholtzstrasse 20, 89081 Ulm}}, \country{Germany}}


\corres{*Matthias Meller. \email{meller.matthias@gmail.com}}


\abstract[]{}

\jnlcitation{\cname{%
\author{Meller M},
\author{Beyersmann J}, and
\author{Rufibach K}} (\cyear{xxxx}),
\ctitle{Joint modelling of progression-free and overall survival and computation of correlation measures}, \cjournal{Stat Med}, \cvol{xxxx;xx:x--x}.}

\maketitle


\section{Estimation of and inference for the correlation coefficient - further simulation scenarios}
\label{sec:EstCorr}

\begin{figure}[h!]
\begin{center}
\setkeys{Gin}{width=1\textwidth}
\includegraphics{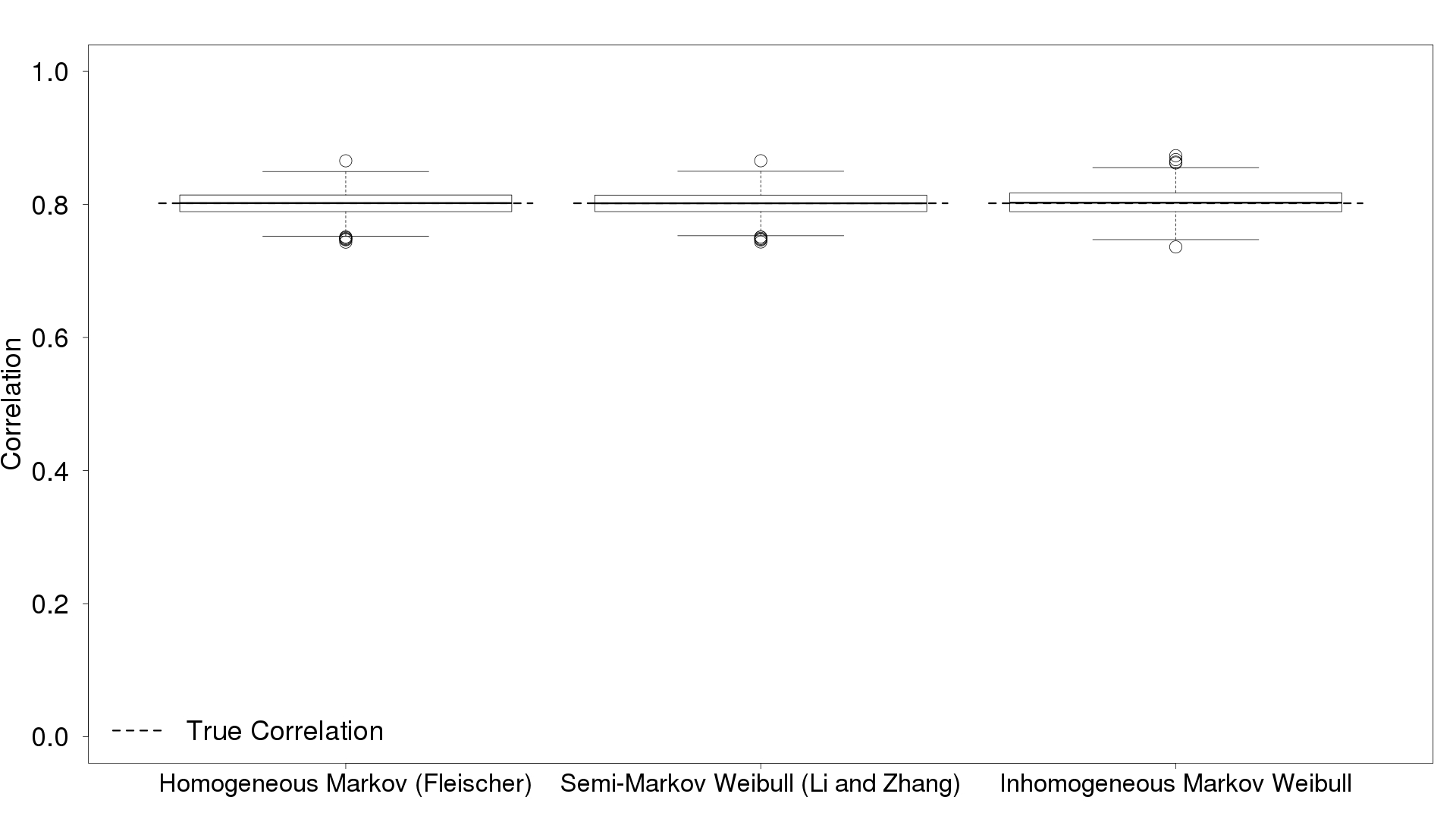}
\end{center}
\caption{Estimated correlation coefficients for time-constant (exponential) transition intensities, (time-)homogeneous Markov process, $1000$ simulation runs.}
\label{fig:corr2}
\end{figure}

\begin{figure}[h!]
\begin{center}
\setkeys{Gin}{width=1\textwidth}
\includegraphics{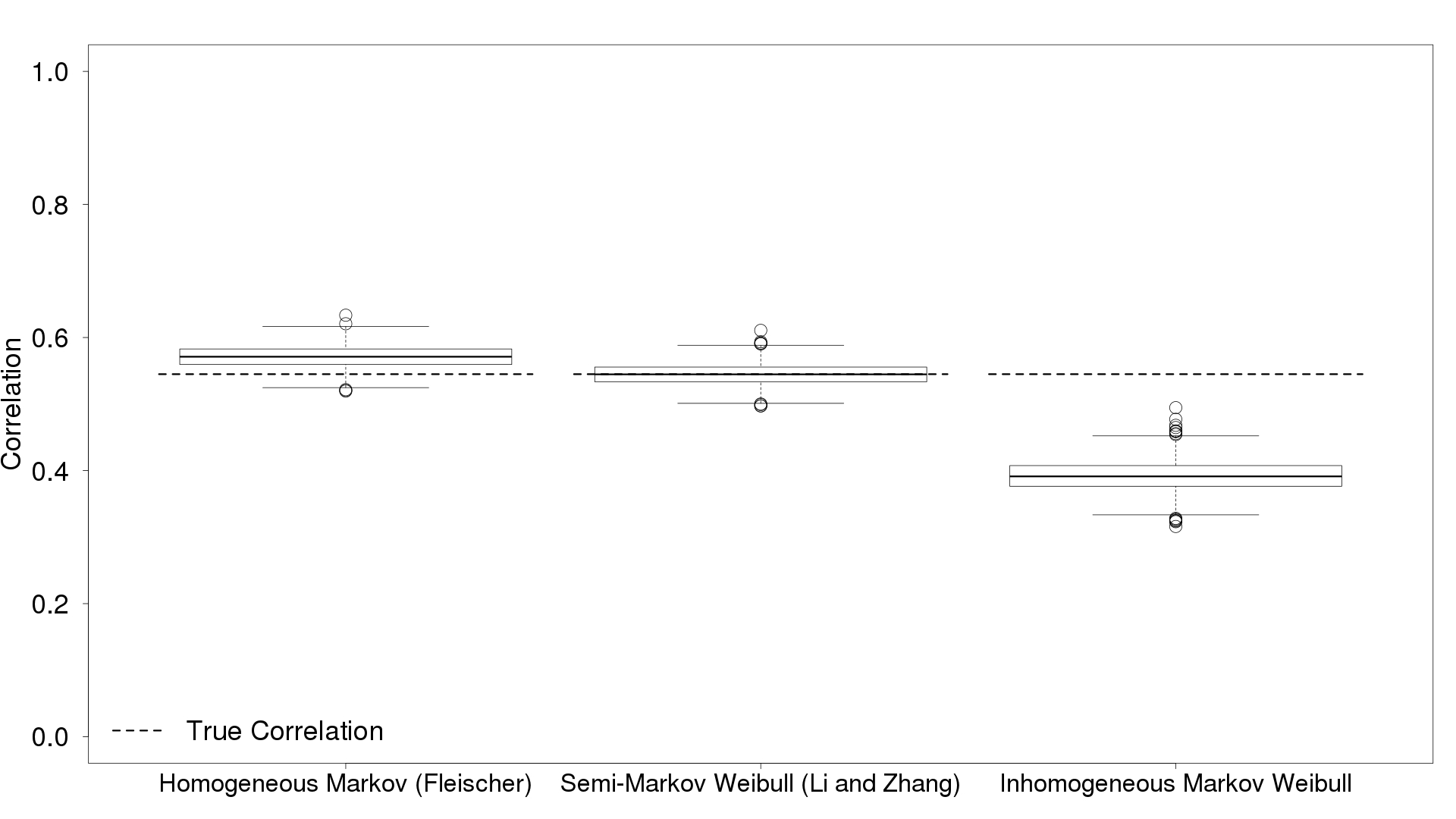}
\end{center}
\caption{Estimated correlation coefficients for Weibull transition intensities (equal shape parameter), semi-Markov process, $1000$ simulation runs.}
\label{fig:corr3}
\end{figure}

\begin{figure}[h!]
\begin{center}
\setkeys{Gin}{width=1\textwidth}
\includegraphics{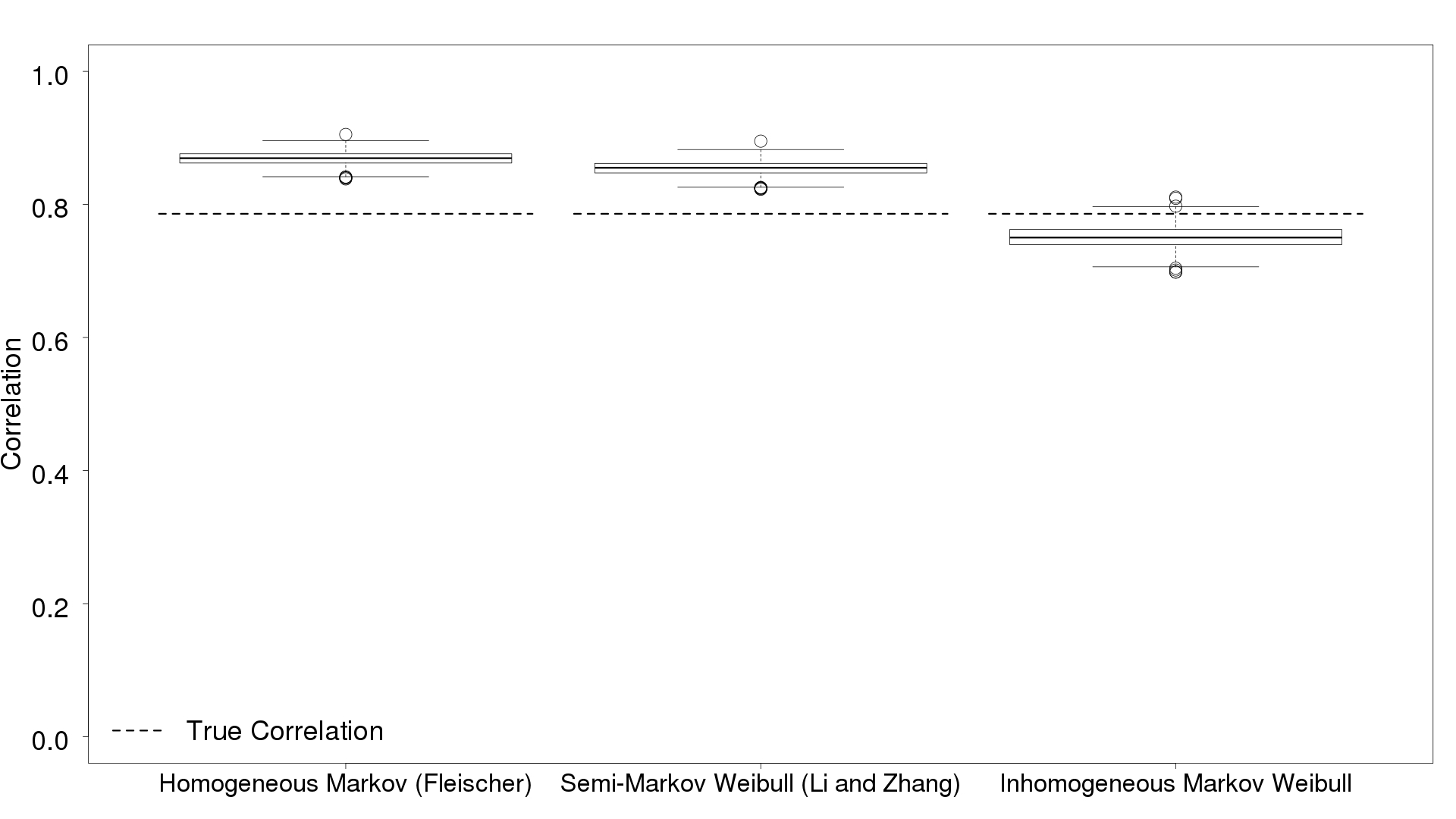}
\end{center}
\caption{Estimated correlation coefficients for Weibull transition intensities (transition-specific shape parameter), semi-Markov process, $1000$ simulation runs.}
\label{fig:corr4}
\end{figure}

\begin{figure}[h!]
\begin{center}
\setkeys{Gin}{width=1\textwidth}
\includegraphics{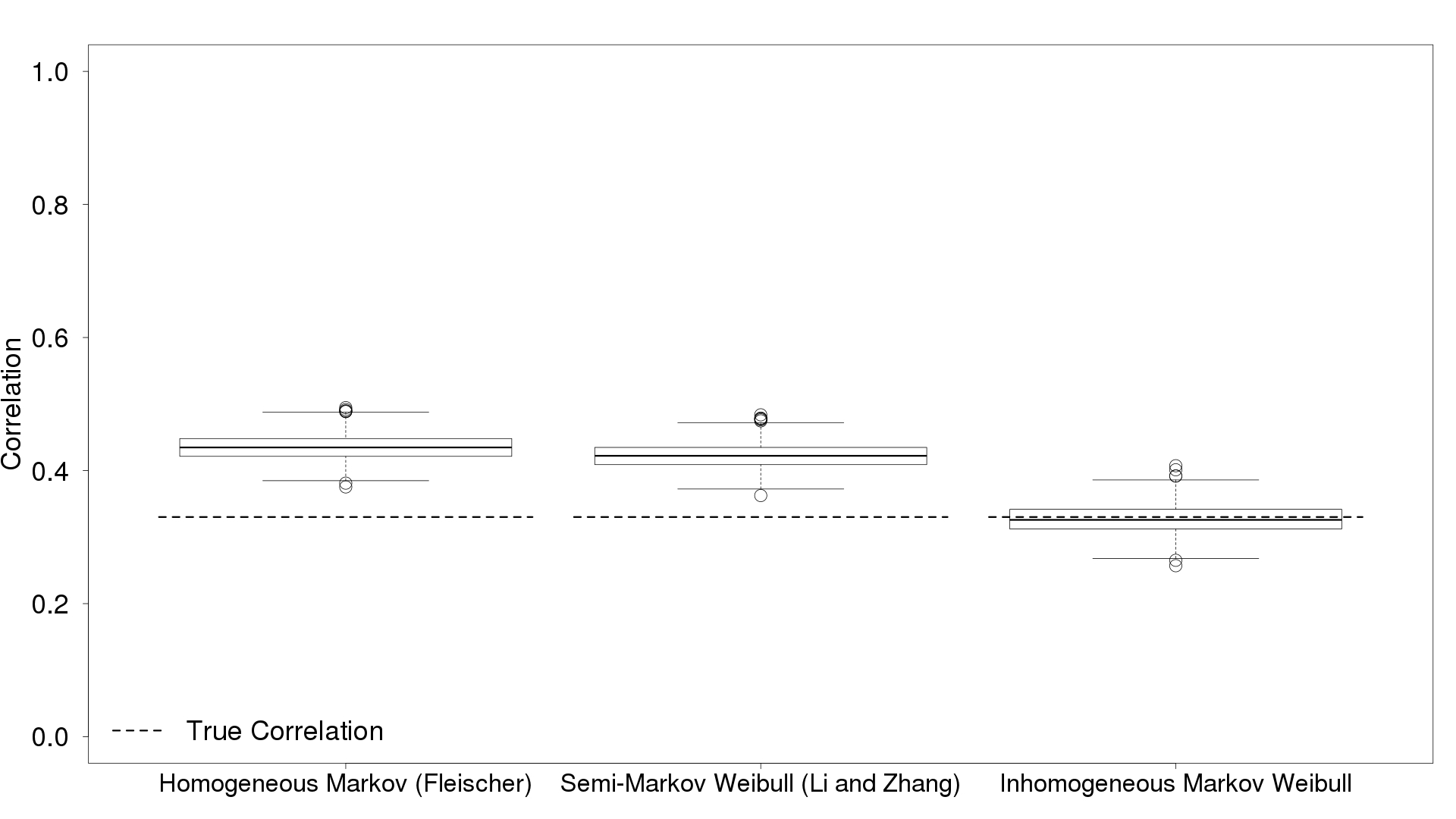}
\end{center}
\caption{Estimated correlation coefficients for Weibull transition intensities (equal shape parameter), time-inhomogeneous Markov process, $1000$ simulation runs.}
\label{fig:corr5}
\end{figure}

Section 5 of Meller et. al \cite{meller_18} presents estimated correlation coefficients from 1000 simulated studies following a time-inhomogeneous Markov model with Weibull transition intensities, where the results using the approaches of \cite{fleischer2009statistical} and \cite{li2015weibull} are compared with the results using a time-inhomogeneous Markov approach based on a counting process likelihood described in Andersen and Keiding \cite{Ande:Keid:mult:2002}.
In Figures ~\ref{fig:corr2}  - ~\ref{fig:corr5}  estimated correlation coefficients from 1000 simulated studies for further four simulation scenarios applying the three different approaches are provided.

Like in Section 5 of Meller et al. \cite{meller_18}, the assumptions in each of the four scenarios for the simulations were as follows: 500 patients were sampled for each scenario and for simplicity reasons no censoring was applied. The underlying model assumption on the illness-death model for each scenario is provided in the first column of Table~\ref{tab:par_scenario} and the distribution and parameter of the transition intensities in the rest of the table.

\renewcommand{\arraystretch}{1.3}
\begin{table}[ht]
\centering
\begin{tabular}{lcccccccc}
\hline
Model Assumption & Distribution & & $\gamma_{01}$ & $\lambda_{01}$ & $\gamma_{02}$ & $\lambda_{02}$ & $\gamma_{12}$ & $\lambda_{12}$ \\
\hline
Homogeneous Markov & Exponential * & & - & 0.6 & - & 0.075 & - & 0.9 \\ 
Semi-Markov & Weibull & & 1.5 & 0.6 & 1.5 & 0.1 & 1.5 & 0.4 \\ 
Semi-Markov & Weibull & &1.8 & 0.3 & 0.5 & 0.05 & 1.2 & 1 \\ 
Time-inhomogeneous Markov & Weibull & &1.3 & 0.85 & 1.3 & 0.1 & 1.3 & 0.3 \\ 
\hline
\end{tabular}
\caption{Model assumption, distribution and true parameter values of transition intensities for each scenario. $\gamma_{ij}$ and $\lambda_{ij}$ correspond to the scale and shape parameter of a Weibull distribution.\\ * Transition intensities in a (time-) homogeneous Markov model are by definition exponentially distributed.}
\label{tab:par_scenario}
\end{table}
\renewcommand{\arraystretch}{1.0}

As expected for a scenario with an underlying time-homogeneous Markov process and resulting time-constant tranisiton intensities, all three approaches, the approaches of Fleischer et al. \cite{fleischer2009statistical}, Li and Zhang \cite{li2015weibull} and Meller et al. \cite{meller_18}, result in unbiased estimation of the correlation between PFS and OS (Figure \ref{fig:corr2}).

Once the illness-death models follows a semi-Markov process, the \textit{Inhomogenous Markov Weibull} approach of Meller et al. \cite{meller_18} results into biased estimation of the correlation coefficient (Figure \ref{fig:corr3} and \ref{fig:corr4}). However, if the semi-Markov assumptions seems plausible in a particular application, the flexibility of the approach would allow for an unbiased estimation of the correlation between PFS and OS by plugging in transition intensites for semi-Markov models shown in the Appendix of Meller et al. \cite{meller_18} into the formulae of Section 2.

Figure \ref{fig:corr5} provides the results of another scenario where the \textit{Inhomogeneous Markov Weibull} approach yields to unbiased estimation of the correlation coefficient. Despite the shared shape parameter of the Weibull transition hazards, the approach of Li and Zhang \cite{li2015weibull} does result into biased estimation and is, in contrary to the approach described in Meller et al. \cite{meller_18},  not adaptable to alternative assumptions.

\section{Real data example - cumulative hazards per treatment arm}

\begin{figure}[h!]
\begin{center}
\setkeys{Gin}{width=1\textwidth}
\includegraphics{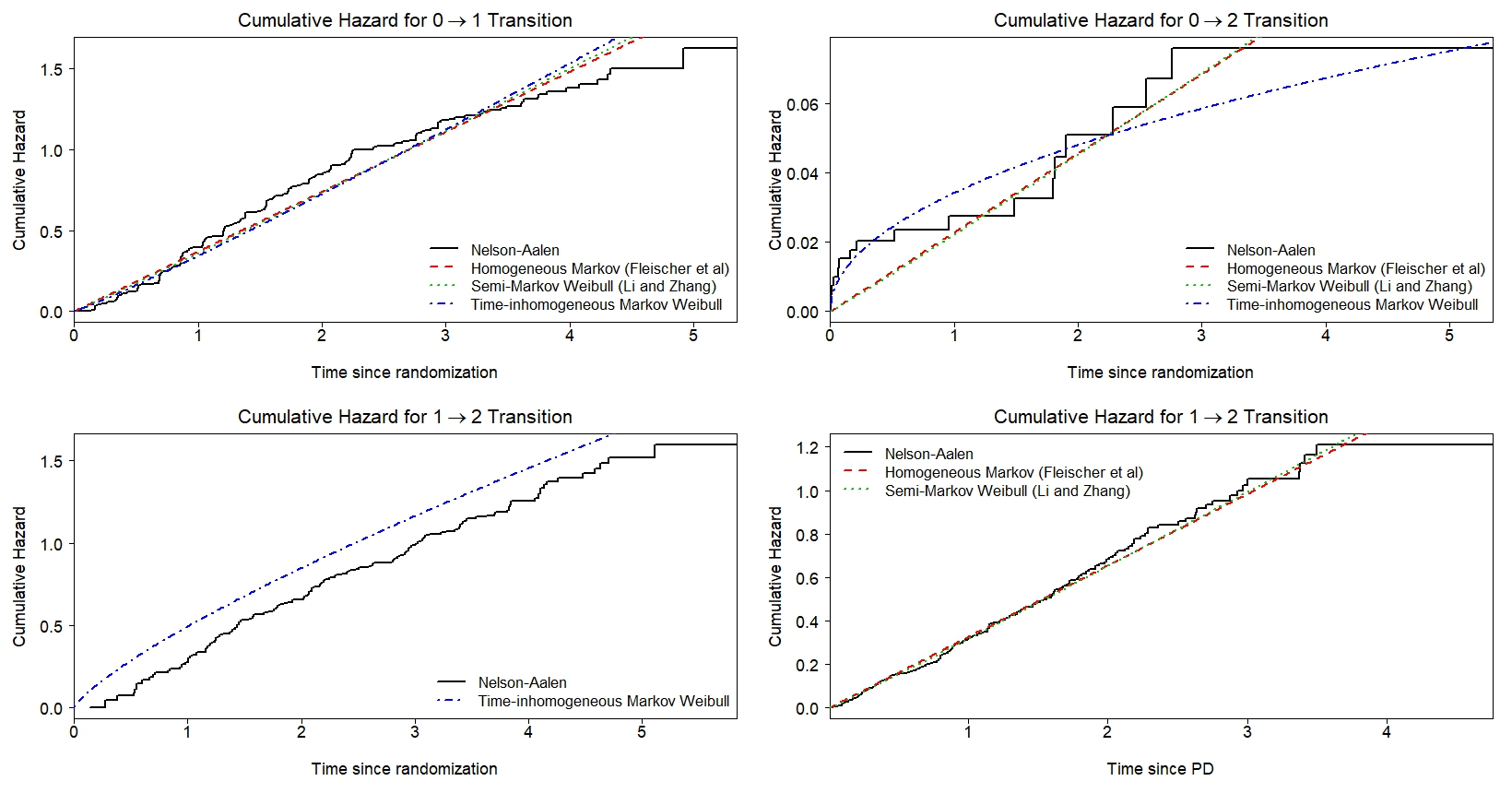}
\end{center}
\caption{CLEOPATRA trial: estimates for cumulative hazards for each transition for the treatment arm.}
\label{fig:CleoCumHaz1}
\end{figure}

\begin{figure}[h!]
\begin{center}
\setkeys{Gin}{width=1\textwidth}
\includegraphics{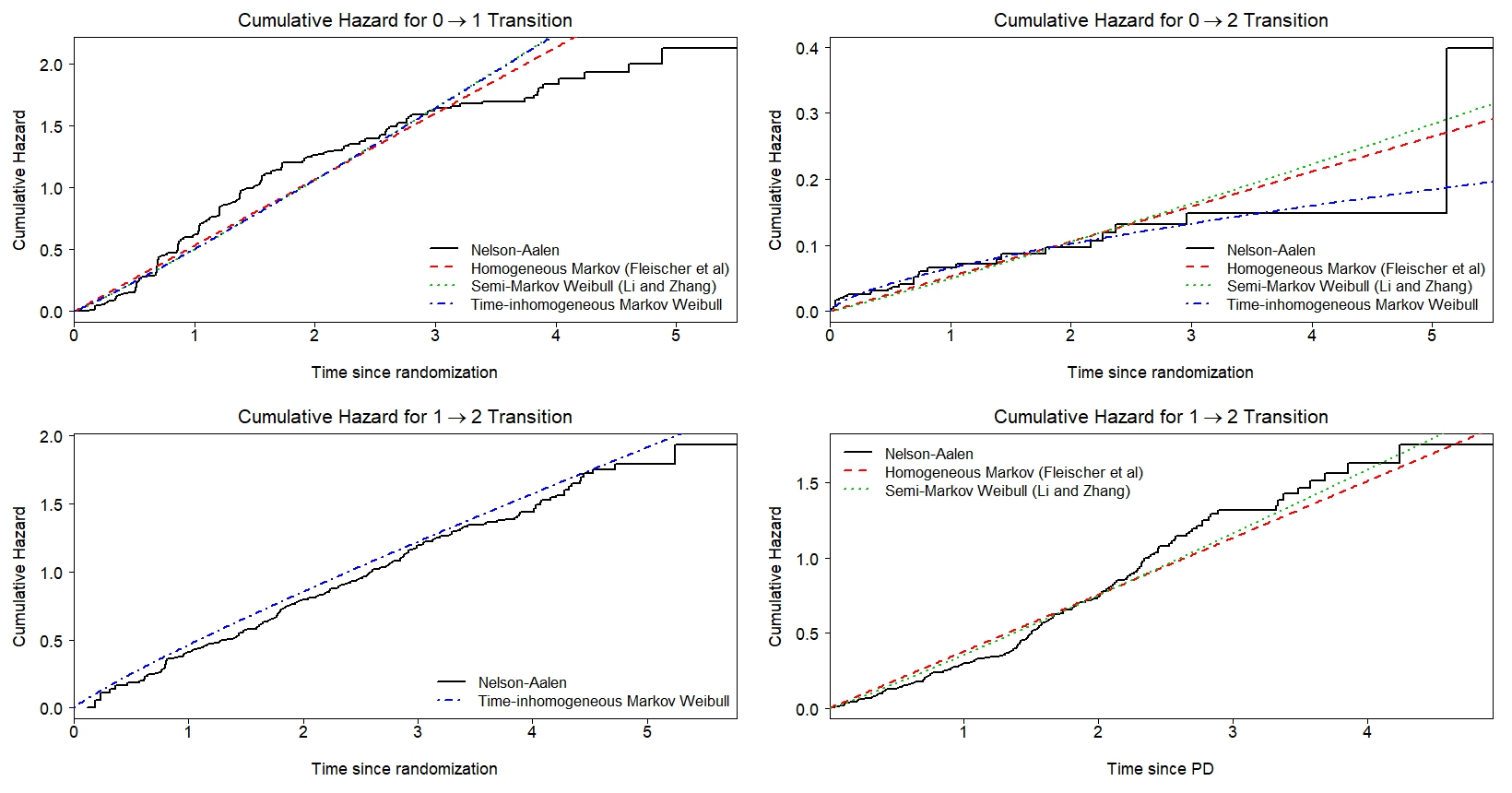}
\end{center}
\caption{CLEOPATRA trial: estimates for cumulative hazards for each transition for the control arm.}
\label{fig:CleoCumHaz2}
\end{figure}

Figure 3 of Meller et al. \cite{meller_18} illustrates estimates of the cumulative hazard function, using the same three approaches applied to the simulated studies in \ref{sec:EstCorr}, for each transition in the CLEOPATRA study \cite{baselga_12} for all patients jointly. Figure \ref{fig:CleoCumHaz1} and Figure \ref{fig:CleoCumHaz2} provide the estimated cumulative incidence functions separated by treatment arm.

\bibliography{literatur}


\clearpage